\documentclass{emulateapj}
\shorttitle{Disk-satellite interaction in disks with density gaps}
\shortauthors{Petrovich \& Rafikov}
\begin{document}

\def\etal{et al.\ \rm}
\def\ba{\begin{eqnarray}}
\def\ea{\end{eqnarray}}
\def\etal{et al.\ \rm}
\def\sgn{\mbox{sgn}}
\def\rmk{\mbox{k}}
\def\Bi{\mbox{Bi}}
\def\Ai{\mbox{Ai}}

\newcommand{\mth}{m_{\rm th}}
\newcommand{\Mth}{M_{\rm th}}
\newcommand{\rs}{r_{\rm s}}
\newcommand{\phip}{\Phi_{\rm p}}
\newcommand{\cs}{c_{\rm s}}
\newcommand{\mplanet}{M_{\rm p}}
\newcommand{\ls}{l_{\rm sh}}
\newcommand{\third}{\rm third}

\title{Disk-satellite interaction in disks with density gaps}

\author{Cristobal Petrovich\altaffilmark{1} \& 
Roman R. Rafikov\altaffilmark{1,2} }
\altaffiltext{1}{Department of Astrophysical Sciences, 
Princeton University, Ivy Lane, Princeton, NJ 08540; 
rrr@astro.princeton.edu}
\altaffiltext{2}{Sloan Fellow}


\begin{abstract}
Gravitational coupling between a gaseous disk and an orbiting 
perturber leads to angular momentum exchange between them which
can result in gap opening by planets in protoplanetary disks and 
clearing of gas by binary supermassive black holes (SMBHs) 
embedded in accretion disks. Understanding the co-evolution 
of the disk and the orbit of the perturber in these circumstances
requires knowledge of the spatial distribution of the torque 
exerted by the latter on a highly nonuniform disk.
Here we explore disk-satellite interaction in disks with 
gaps in linear approximation both in Fourier and in physical space, 
explicitly incorporating the disk non-uniformity in the fluid equations.
Density gradients strongly displace the positions of Lindblad 
resonances in the disk (which often occur at multiple locations),
and the waveforms of modes excited close to the gap edge get modified
compared to the uniform disk case. The spatial distribution of the 
excitation torque density 
is found to be quite different from the existing prescriptions:
most of the torque is exerted in a rather narrow region near 
the gap edge where Lindblad resonances accumulate,
followed by an exponential fall-off with the distance from 
the perturber. Despite these differences, for a given gap profile
 the full integrated torque exerted on the disk agrees with the conventional 
 uniform disk theory prediction at the level of $\sim10\%$. 
 The nonlinearity of the density wave 
excited by the perturber is shown to decrease as the wave travels
out of the gap, slowing down its nonlinear evolution and damping.
Our results suggest that gap opening in protoplanetary disks 
and gas clearing around SMBH binaries can be more efficient than the 
existing theories predict. They pave the way for self-consistent 
calculations of the gap structure and the orbital evolution of the
perturber using accurate prescription for the torque density
behavior.
\end{abstract}



\section{Introduction.}  
\label{sect:intro}


Gravitational coupling between a gaseous disk and an external 
orbiting body plays an important role in many astrophysical systems, 
including disk-planet interaction in protoplanetary disks, 
orbital evolution of supermassive black hole (SMBH) binaries 
surrounded by gaseous disks, dynamics of accretion disks 
in cataclysmic variables, and so on. This interaction leads 
to angular momentum exchange between the disk and the perturber,
causing the orbit of the latter to evolve --- an effect 
responsible for the planet migration 
in protoplanetary disks \citep{ward86,ward97} and the black hole 
inspiral in the case of SMBH binaries embedded in gaseous disks 
\citep{ivanov99,gould00}

Tidal disk-satellite\footnote{We will refer as 
disk-satellite or disk-planet to any type 
of system where the tidal coupling is present, e.g., an 
SMBH binary surrounded by the disk, an accreting 
white dwarf-main sequence star binary, etc.} 
coupling also modifies the distribution 
of the gas surface density and can result in gap opening in 
protoplanetary disks \citep{ward97} and cavity formation 
around the SMBH binaries \citep{armitage02} if the perturber 
is massive enough. Gap opening necessarily weakens the tidal 
coupling between the perturber and the disk, and
slows the orbital migration of the 
perturber, switching it to the so-called Type II regime, 
\citep{ward86}. 
Thus, gap opening can be an important ``parking mechanism'' 
preventing massive planets from migrating all the way to 
the central star. 

In steady state the gap structure is determined by the local 
balance between the planetary torque acting on the disk and the
divergence of the external stress $T_{r\varphi}$, which can be 
due to magnetorotational instability or some other mechanism of 
intrinsic angular momentum transport in the disk:
\ba
\frac{dT}{dr}\Bigg|_d=\frac{dT_{r\varphi}}{dr}.
\label{eq:gap_balance}
\ea
Here $dT/dr|_d$ is the torque density (the amount of torque
per unit radial distance) {\it deposited in the disk} by the 
density waves originally launched by the planetary tide closer 
to the planet. The deposition of the wave angular momentum
in the disk occurs by virtue of some linear \citep{takeuchi96} 
or nonlinear \citep{GR01} {\it damping} process. In general 
the spatial distribution of $dT/dr|_d$ can be described as
\ba
\frac{dT}{dr}\Bigg|_d=\mathcal{L}*\left(\frac{dT}{dr}\right),
\label{eq:2pieces}
\ea
where the intrinsically non-local operator $\mathcal{L}$ 
describes the wave damping, and $dT/dr$ is the {\it excitation}
torque density --- the rate at which planetary potential adds 
angular momentum to the propagating density waves per unit 
radial distance. 

Thus, in order to understand the details of gap opening 
one must understand both the \textit{excitation} of waves and 
their \textit{damping}, a key point first highlighted by 
\citet{lunine82}, \citet{greenberg83}, and \citet{GN89}. 
Once the description of the wave damping (i.e. the explicit form 
of the operator $\mathcal{L}$) is available, the steady state 
gap structure is fully determined by the spatial structure of 
the excitation torque density $dT/dr$. 

In this work, following the approach of 
Rafikov \& Petrovich (2012; hereafter RP12) we 
concentrate only on the process of wave excitation, 
which at least in some cases \citep{GR01} can be studied 
separately from the wave dissipation (when the latter is weak).
Our goal is to provide a 
self-consistent calculation of the excitation torque 
density $dT/dr$ in {\it non-uniform} disks affected by 
gap formation. Prior to this work the majority of existing 
studies of the gap and cavity opening in disks have 
adopted the following simple prescription for the
excitation torque density behavior in non-uniform disks: 
\ba
\frac{dT(r)}{dr}=\frac{\Sigma(r)}{\Sigma_0}
\frac{dT(r)}{dr}\Bigg|_{u},
\label{eq:dTdx_nu}
\ea
where $dT/dr|_u$ is the excitation torque density in a 
{\it uniform} disk with surface density $\Sigma_0$, and 
$\Sigma(r)$ is the spatially varying surface density in a 
non-uniform disk. 

The behavior of $dT/dr|_u$ has been previously derived from 
direct numerical simulations
\citep{bate03,valborro,dangelo08,dangelo10,dong11a,dong11b}
and from analytical linear studies
(GT80; Muto \& Inutsuka 2009; RP12). 
In particular, in their pioneering study of disk-satellite 
interaction \citet{GT80} have shown that far from the 
perturber, at radial separations from its orbit 
$|r-r_p|$ ($r_p$ is the semimajor axis of the circular orbit of 
the planet) exceeding the disk scale height $h$, the 
uniform disk excitation torque density\footnote{We assume that 
GT80 calculation refers to $dT/dr$ and not to $dT/dr|_d$ 
since no explicit dissipation mechanism was 
mentioned in their work.} $dT/dr|_u$ is given by
\ba
\frac{dT}{dr}\Bigg|_u  &\to& \mbox{sign}(r - r_p) C_{ GT80}
\frac{(GM_p)^2\Sigma_0}{\Omega^2 |r - r_p |^4},\nonumber \\
C_{\rm GT80} &=&
\frac{32}{81}\left[2K_0\left(\frac{2}{3}\right)+
K_1\left(\frac{2}{3}\right)\right]^2\approx 2.5,
\label{eq:dTdx_GT}
\ea
where $M_p$ is the planetary mass, 
$\Sigma_0$ is the disk surface density assumed
to be uniform on scales $\sim|r - r_p|$, 
$\Omega$ is the angular frequency of the disk 
at $r_p$, and $K_n$ is the modified Bessel function of order $n$.
Note that strictly speaking $dT/dr$ in equation (4) represents the
 {\it force density} and not the angular momentum density 
 as it lacks an additional factor of $r_p$. 
This is because our subsequent calculations
are cast in the shearing sheet framework in which the concept
of $r_p$ is not well defined.

The general scaling $dT/dr|_u\propto |r-r_p|^{-4}$ has
been also confirmed by \citet{LP79} using impulse approximation.
A number of studies of gap opening by planets 
\citep{LP86,trilling98,bryden99,armitage02,varni04,crida06}
and orbital evolution of SMBH binaries surrounded by gaseous 
disks \citep{gould00,armi-nata02,lodato09,alexander11}
have used the prescription\footnote{It should also be pointed 
out that in all these studies the excitation torque density 
$dT/dr$ was identified with the deposition torque density 
$dT/dr|_d$, thus ignoring the process of wave dissipation 
altogether, see equation (\ref{eq:2pieces}).} 
(\ref{eq:dTdx_nu}) combined with 
asymptotic GT80 formula (\ref{eq:dTdx_GT}) for $dT/dr|_u$, 
in some cases with a value of the constant pre-factor 
different from $C_{\rm GT80}$ \citep{LP84,armi-nata02}. 

Recently, \citet{dong11a} carried out high-resolution
numerical simulations of disk-planet interaction in the
two-dimensional shearing sheet geometry. Contrary 
to the prediction (\ref{eq:dTdx_GT}) of GT80 they found 
that the one-sided excitation torque density $dT/dr|_u$ 
does not maintain a fixed sign but 
rather changes from positive to negative at a distance 
of $\approx 3.2h$ from the perturber --- a result recently 
confirmed by global simulations of \citet{duffell}. 
This negative torque density phenomenon 
was then explained analytically in the framework of 
linear theory by RP12, who traced its origin to the 
overlap of Lindblad resonances in the vicinity of the 
perturber's orbit (which was ignored in GT80).

This unexpected finding casts serious doubt on the validity 
of simple prescription (\ref{eq:dTdx_nu}) in non-uniform disks.
Indeed, directly substituting $dT/dr|_u$ computed by RP12 
in equation (\ref{eq:dTdx_nu}) can result in {\it negative} total
torque exerted on the disk by the perturber if the gap is 
sufficiently wide and deep (in view of RP12 results this 
is certainly true in the extreme case of the gas density 
being exactly zero at separations less than $3.2h$ from 
the planetary orbit). This conclusion is unphysical (planet 
would not open a gap in the first place), suggesting that the 
oversimplified prescription (\ref{eq:dTdx_nu}) incorporating 
the disk non-uniformity only through the direct multiplication 
by the local surface density (so that density {\it gradients} 
do not affect disk-satellite coupling) needs to be revised. 

In this work, we provide a fully self-consistent linear calculation
of the disk-satellite interaction by explicitly incorporating the 
disk non-uniformity in the fluid equations and properly accounting 
for the effect of density gradients on the waveforms of perturbed 
fluid variables. In this way we are able to compute the torque 
density and angular momentum flux in Fourier and physical space 
and demonstrate significant differences with the results obtained 
using the prescription (\ref{eq:dTdx_nu}), especially at 
the gap edges where the density gradients are large.

Our paper is structured as follows. In \S \ref{sect:setup} 
we describe the problem setup, in particular the governing 
equations, the assumed gap density profile, and the Lindblad 
resonances in non-uniform disks. Our numerical procedure and
the results for the density wave behavior are described in 
\S \ref{sect:results}. Calculations of the torque exerted
by the perturber on the disk in Fourier and real space are 
described in \S \ref{sect:torque_fourier} and 
\ref{sect:torque_physical}, respectively. Finally, we discuss 
our results, including their astrophysical applications, 
in \S \ref{sec:discussion}.


\section{Problem setup.}  
\label{sect:setup}


We start by deriving a system of linearized equations and 
describing the adopted underlying surface density distribution 
--- the ingredients needed to 
calculate the spatial behavior of the perturbed quantities.


\subsection{Basic equations.}  
\label{subsect:eqs}

We study the tidal coupling of a planet with a non-uniform disk 
in the shearing sheet geometry \citep{GL65}, which allows us to 
neglect geometric curvature effects while preserving the
main qualitative features of the system. We also neglect the 
vertical dimension and assume the disk to be two-dimensional. 
The dynamics of fluid is then governed by the following 
equations\footnote{Analogous equation (3) in RP12 has a typo:
the term $4A\Omega z{\bf e}_x$ should be $4A\Omega x{\bf e}_x$. 
This typo does not affect any other part of the paper.}
(e.g. Narayan \etal 1987; hereafter NGG):
\ba
 \frac{\partial {\bf v}}{\partial t}+
\left({\bf v}\cdot\nabla\right){\bf v}+ 
2{\bf \Omega}\times {\bf v} +4A\Omega x{\bf e}_x
&=&-\frac{\nabla P}{\Sigma}-\nabla\Phi_p,
\label{eq:motion}\\
 \frac{\partial \Sigma}{\partial t}+\nabla\cdot
\left({\bf v}\Sigma\right)&=&0,
\label{eq:cont}
\ea
where ${\bf v}=(u,v)$ is the fluid velocity with components in
the $x\equiv r-r_p$ (radial) and $y$ (azimuthal) directions correspondingly, 
$P$ is gas pressure, $\Phi_p$ is the planetary potential, 
${\bf \Omega}_p\equiv\Omega_p{\bf e}_z$ is the Keplerian rotation 
frequency at the planet position ($r=r_p$; we will subsequently
drop the subscript ``p'' for brevity),
and $A\equiv (r/2)(d\Omega/dr)$ is the shear rate at the same location. 
Additionally, in our notation $c_s$ is the sound speed and $B=\Omega+A$
is Oort's B constant. 
 
For an isothermal disk with no planet present ($\phi\equiv0$)
and for an arbitrary background density profile
$ \Sigma = \Sigma_{0}(x)$, the Equations (\ref{eq:motion}) and (\ref{eq:cont})
have the following exact steady-state solution 
 \begin{eqnarray}
u_{0}=0, \quad v_{0}=2Ax+\frac{c_s^{2}}{2\Omega} \frac{\partial \ln \Sigma_{0}}{\partial x}.
\label{eq:v_0}
\end{eqnarray}

After having specified our steady-state solution $\{\Sigma_0,u_0,v_0\}$, we
introduce the planetary potential as a perturbation and study the 
linear behavior of the perturbed density $\Sigma_1$ and velocity field 
$\delta \vec{v}=(u_1,v_1)$.
By ignoring the quadratic perturbed quantities that appear 
in the Equations (\ref{eq:motion}) and (\ref{eq:cont})
one can get the following general system
\begin{eqnarray}
&&\frac{\partial u_1}{\partial t}+\left(2Ax+\frac{c_s^{2}}{2\Omega}
 \frac{\partial \ln \Sigma_{0}}{\partial x}\right)\frac{\partial u_1}{\partial y}
-2\Omega v_1 +
\nonumber \\
&& ~~~~~~~~~~~~~~~~~~~~~~~~~~~~~~~~~~~~~
 c_s^{2}\frac{\partial\left(\Sigma_1/\Sigma_0\right)}{\partial x}=- \frac{\partial \Phi_p}{\partial x}\\
&&\frac{\partial v_1}{\partial t}+\left(2Ax+\frac{c_s^{2}}{2\Omega} 
\frac{\partial \ln \Sigma_{0}}{\partial x}\right)\frac{\partial v_1}{\partial y}+
 \nonumber \\
&& 
	+\left(2B+\frac{c_s^{2}}{2\Omega} \frac{\partial^{2} \ln \Sigma_{0}}{\partial x^{2}} \right) u_1 + 
c_s^{2}\frac{\partial \left(\Sigma_1/\Sigma_0\right)}{\partial y} =-\frac{\partial \Phi_p}{\partial y}\\
&&\frac{\partial\left(\Sigma_1/\Sigma_0\right)}{\partial t}+
	\left(2Ax+\frac{c_s^{2}}{2\Omega} \frac{\partial \ln \Sigma_{0}}
	{\partial x}\right)\frac{\partial\left(\Sigma_1/\Sigma_0\right)}{\partial y}+ \nonumber \\
&&~~~~~~~~~~~~~~~~~~~~
+ u_1\frac{\partial \ln\Sigma_{0}}{\partial x}+
\frac{\partial u_1}{\partial x}+\frac{\partial v_1}{\partial y}=0
\end{eqnarray}

We then represent all perturbed fluid variables via
Fourier integrals as 
$\{\Sigma_1,u_1,v_1,\Phi_p \}=(2\pi)^{-1}\int_{\infty}^{\infty} 
\exp(ik_yy)\{\delta\Sigma,u,v,\phi\}$, which takes care
of the $y$-dependence leaving us with the following set of equations
in the $x$-coordinate only.
Thus, by defining
\begin{eqnarray}
\sigma(x)&=&-2Ak_{y}x-\frac{c_s^{2}k_{y}}{2\Omega} \frac{\partial \ln \Sigma_{0}}{\partial x}, 
\label{eq:sigma} \\
\tilde{B} &=&B +\frac{c_s^{2}}{4\Omega} \frac{\partial^{2} \ln \Sigma_{0}}{\partial x^{2}}
\label{eq:tildeB} 
\end{eqnarray}
we get the linearized system in steady state 
($\partial /\partial t \equiv 0$)
\begin{eqnarray}
-i\sigma u-2\Omega v +c_s^{2}\frac{\partial \left(\delta\Sigma/\Sigma_0\right)}{\partial x}
		&=&- \frac{\partial \phi}{\partial x}
		\label{eq:system_v_1}\\
-i\sigma v+2 \tilde{B} u +i k_{y}c_s^{2}\frac{\delta\Sigma}{\Sigma_{0}} &=& - ik_{y} \phi
		\label{eq:system_v_2}\\
-i\sigma \frac{\delta\Sigma}{\Sigma_{0}}+u\frac{\partial \ln \Sigma_{0}}{\partial x}+
\frac{\partial u}{\partial x}+ik_{y}v&=&0,
		\label{eq:system_v_3}
\end{eqnarray}
where the Fourier component of the planetary potential produced
by a point mass at the origin and its derivative are given by
\ba
\phi(k_y,x)&=&-\frac{GM_p}{\pi}K_0\left(|k_yx|\right), 
\label{eq:pot}\\
\frac{\partial \phi}{\partial x}&=&\sgn(x)\frac{GM_p}{\pi}k_y 
K_1\left(|k_yx|\right).
\label{eq:pot_der}
\ea

Solving (\ref{eq:system_v_1})-(\ref{eq:system_v_3}) for the azimuthal
velocity perturbation $v$, we obtain
the second order ordinary  differential equation
\begin{eqnarray}
&&\frac{\partial^{2} v}{\partial x^{2}}+
\frac{\partial v}{\partial x}\frac{\partial\ln \left(\Sigma_{0}/\Delta\right)}{\partial x}+   
\nonumber \\
&&+v\left[
\frac{\sigma^{2}}{c_s^{2}}-\frac{\Omega}{c_s^{2}\tilde{B}}\Delta +
\frac{k_{y}\sigma \Lambda}{2\tilde{B}}
\frac{\partial\ln\left(\sigma \Lambda/\Delta \right)}{\partial x}\right] = 
\nonumber\\
&&-\frac{\partial \phi}{\partial x}
\frac{2\tilde{B}}{c_s^{2}}
+\phi \left[ 
\frac{k_{y}\sigma}{c_s^{2}} +\frac{k_{y}^{2}\Lambda}{2\tilde{B}}\frac{\partial\ln\left(\Lambda/\Delta\right)}{\partial x}\right],
\label{eq:v_eq}
\end{eqnarray}
where we have defined
\ba
\Lambda&=& 1-\frac{2}{k_{y}\sigma}\left(\frac{\partial \tilde{B}}{\partial x}-
\tilde{B}\frac{\partial\ln \Sigma_{0}}{\partial x}\right), 
\label{eq:Lambda}\\
\Delta&=&c_s^2k_{y}^{2}\Lambda+4\tilde{B}^{2}.
\label{eq:Delta}
\ea

The perturbed radial velocity $u$ and the surface density 
perturbation $\delta\Sigma$ are directly expressed in terms of 
$v$ via the following relations:
\ba
&& u=-\frac{i}{\Delta}\left(k_y c_s^2\frac{\partial v}
{\partial x}-2 \tilde{B} \sigma v + 2\tilde{B}k_y\phi\right),
\label{eq:u}\\
&& \delta\Sigma=\frac{\Sigma_0}{\Delta}\left(2\tilde{B}\frac{\partial v}
{\partial x} +k_y\sigma\Lambda  v - k_y^2 \Lambda\phi\right).
\label{eq:dS}
\ea

One can easily check that in the limit 
$\partial \Sigma_{0}/\partial x\to 0$ equation (\ref{eq:v_eq}) 
reduces to equation (5) of RP12 
which was derived for a uniform disk.


\subsection{Disk models}
\label{sect:models}

\begin{figure}
\centering
\includegraphics[width=8.7cm,height=6.5cm]{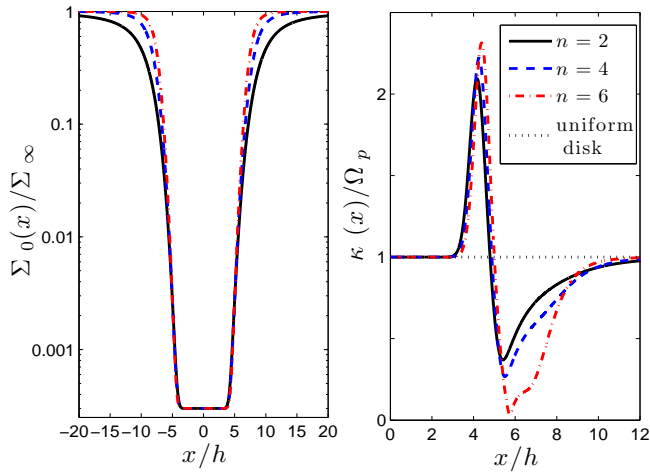}
\caption{Density profiles (left panel) and 
epicyclic frequencies $\kappa$ (right panel) 
for the gap models described by Eq. (\ref{eq:sigma_1}) 
with parameters 
$\delta=5h$, $\Sigma_{\rm min}=3\times10^{-4}$, and $n=2,4,6$.
\label{fig:sigma_profiles}
}
\end{figure}

We model the density structure of an axisymmetric gap around 
the satellite orbit by a particular three parameter density 
profile (symmetric with respect to $x=0$)
\begin{eqnarray}
\frac{\Sigma_{0}(x)}{\Sigma_\infty}=1-\frac{1-\Sigma_{\rm min}}
{1+(x/\delta)^{n} \exp
\left[-\left( 2.5\delta/x\right)^{2}\right]},
\label{eq:sigma_1}
\end{eqnarray}
where the surface density is normalized to its value at 
infinity $\Sigma_\infty$. In this profile the dimensionless parameter 
$\Sigma_{\rm min}$ controls the depth of the gap, $\delta$ 
regulates its width, while an even positive integer $n$ sets 
the steepness of the density gradient at the gap edge. The 
density variation between $\Sigma_{\rm min}\Sigma_\infty$
inside the gap and $\Sigma_\infty$ outside of it 
is predominantly due to the power law dependence on $x$ in the 
denominator of the second term in the right hand side of 
the equation (\ref{eq:sigma_1}).
The exponential dependence in the denominator is introduced
only to ensure a flat bottom at $x\lesssim \delta$ and allows us to 
avoid large density gradients close to the origin. This simplifies 
our subsequent analysis and helps us get important contributions 
to the torque from just outside the gap, exactly where we want 
to test our theory. The factor 2.5 in the exponential is introduced 
for the parameter $\delta$ to better correspond to the width of the
inner, flat part of the gap.

In the left panel of Figure \ref{fig:sigma_profiles} we 
plot the density profile for the disk model with parameters
$\delta=5h$, $\Sigma_{\rm min}=3\times10^{-4}$,
and with $n=2,4,6$, respectively. These profiles
represent rather broad gaps: for the models with $\delta=5h$ 
shown in Figure \ref{fig:sigma_profiles} one finds that 
the density $\Sigma_{\infty}/2$ is reached at $|x|\approx10h$, 
$|x|\approx8.5h$, and $|x|\approx7.7h$ for $n=2$, 4, and 6 
respectively. Evidently, as $n$ is increased the profile becomes  
boxier and the gradients at the gap edge (they are largest for  
$4h\lesssim |x| \lesssim 15 h$) grow.


\subsection{Shifted Lindblad resonances}
\label{sect:lindblad}

Non-zero pressure gradients intrinsic to inhomogeneous disks
affect the unperturbed azimuthal velocity 
(\ref{eq:v_0}) and change the locations of the Lindblad 
resonances, at which different harmonics of planetary potential 
couple to the disk, compared to the purely Keplerian 
case. In Appendix A we demonstrate that in the shearing sheet 
approximation a given $k_y$ harmonic has Lindblad resonance
condition satisfied at $x_L(k_y)$ satisfying the following 
implicit relation: 
 \begin{eqnarray}
k_y(x_L)=\pm\frac{2}{3}\frac{\kappa(x_L)}{\Omega_p}
\left(x_L-\frac{h^2}{3}\frac{\partial \ln \Sigma_{0}}
{\partial x}\Big|_{x_L}\right)^{-1},
\label{eq:k_y}
\end{eqnarray}
where the modified value of the local epicyclic frequency 
$\kappa$ for the inhomogeneous disk is given by (see Appendix A)
\ba
\kappa(x)=\Omega_p\left(1+ h^2\frac{\partial^{2} \ln \Sigma_{0}}
{\partial x^{2}}\right)^{1/2}.
\label{eq:kappa}
\ea
In a uniform disk ($\partial \ln\Sigma_0/\partial x=0$) one 
recovers the usual expressions $\kappa=\Omega_p$ and 
$x_L(k_y)=2/(3k_y)$.

\begin{figure}
\centering
\includegraphics[width=8.5cm,height=7cm]{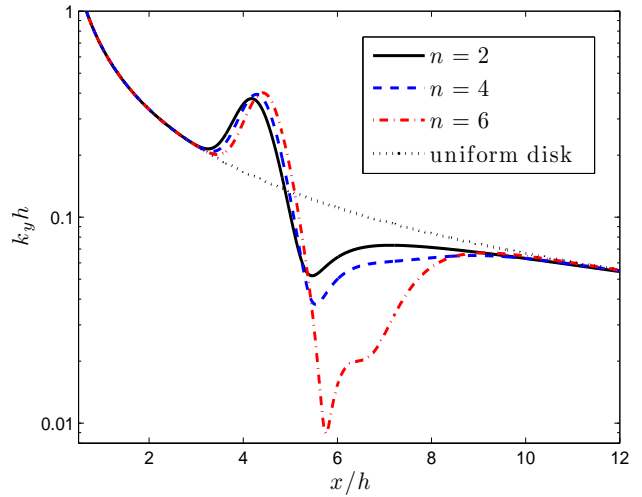}
\caption{Azimuthal wavenumber $k_y$ as a function of the 
Lindblad resonance position given by Eq. (\ref{eq:k_y}) for the
surface profile described by Eq. (\ref{eq:sigma_1}) with parameters 
$\delta=5h$, $\Sigma_{\rm min}=3\times10^{-4}$, and $n=2,4,6$. 
The dotted line indicates the uniform disk
case $k_y=2/(3x)$ for comparison.
\label{fig:lindblad}
}
\end{figure}

For a class of density profiles (\ref{eq:sigma_1}) considered
in this work  the modification to the $x_L(k_y)$
relation (\ref{eq:k_y}) compared to the purely Keplerian 
case comes predominantly 
from the change of the local epicyclic frequency
as given by equation (\ref{eq:kappa}). For this reason we 
explore the dependence of $\kappa$ on $x$ in some detail and plot 
it in the right panel of Figure \ref{fig:sigma_profiles} for
the density profiles shown in the left panel of the same Figure.

Deep inside the gap $\Sigma_0$ is almost constant because of the
assumed exponential dependence in equation (\ref{eq:sigma_1}) and
$\kappa(x)\to \Omega_p$. As the gap edge is approached, at 
$|x| \gtrsim 4h$, $\partial^2\ln\Sigma_0/\partial x^2$ increases 
and $\kappa$ does the same: in all displayed models 
$\kappa$ goes up to $\gtrsim 2\Omega_p$. Beyond $|x| \gtrsim 5h$
the density profiles change their curvature from positive to negative 
and $\kappa$ becomes smaller than $\Omega_p$. At even larger 
$|x|$ the density profile flattens out and $\kappa$ converges 
to $\Omega_p$ at $|x| \gtrsim 15h$. The deviations of 
the epicyclic frequency compared to its value in a uniform disk 
become more extreme as we increase $n$ (increase boxiness of
$\Sigma_0(x)$), with the differences between models 
being more dramatic in regions where $\kappa<\Omega_p$. 
The minimum values of $\kappa$ are $0.36\Omega_p$, 
$0.27\Omega_p$, and $0.03\Omega_p$ for profiles with $n=2$, 4, 
and 6, respectively, see Figure \ref{fig:sigma_profiles}.

In fact, one can see
from equation (\ref{eq:kappa}) that depending on the curvature of 
$\Sigma_0$ the gap profile could have $\kappa^2<0$ at the gap edge,  
which would mean that the disk is Rayleigh unstable. For this 
reason in this work we ignore profiles with extremely deep gaps 
and strong density gradients at their edges since these are likely 
to result in $\kappa^2<0$ at some place. Our boxiest model with 
$n=6$ might be regarded as a limiting profile (for assumed values 
of $\delta$ and $\Sigma_{\rm min}$) since the epicyclic frequency 
closely approaches zero at $x\approx 6h$.

Armed with this knowledge we plot in Figure 
\ref{fig:lindblad} the dependence (\ref{eq:k_y}) for the 
density model given by equation (\ref{eq:sigma_1}) with
the same set of parameters as in Figure \ref{fig:sigma_profiles}: 
$\delta=5h$, $\Sigma_{\rm min}=3\times10^{-4}$, and $n=2,4,6$.
By comparing these two Figures one can see that the resonances 
are displaced {\it outwards} relative to the uniform density case in 
regions where $\kappa>\Omega_p$, while the opposite 
happens when $\kappa<\Omega_p$. As a result, there is a 
{\it concentration of the resonances} at the gap edge where this 
transition occurs. 

Indeed, in a uniform disk with $\kappa=\Omega_p$ the radial 
interval $4h<|x|<6h$ contains Lindblad resonances only for the 
modes with $k_yh$ in the range $(0.11,0.17)$. At the same time,
for the gap in the form (\ref{eq:sigma_1}) with $n=2$ the 
same radial interval contains Lindblad resonances for modes 
satisfying $0.05 \lesssim k_yh \lesssim 0.37$; for $n=6$ model 
modes with $0.009 \lesssim k_yh \lesssim 0.4$ are being excited 
in the same radial interval. Thus, concentration of resonances 
at the gap edge is more pronounced for density profiles with 
larger gradients meaning larger deviations of $\kappa$ from 
$\Omega_p$.

Another striking feature of the $x_L(k_y)$ dependence clearly 
visible in Figure \ref{fig:lindblad} is that even for not 
very sharp gaps (for rather low values of $n$) there are 
regions close to the gap edge
where a single $k_y$ mode can be excited at two or three 
different locations (for a fixed sign of $x$), as happens 
e.g. for $k_yh=0.3$ in this Figure. This splitting of 
resonances in inhomogeneous disks has important implications 
for the torque behavior in Fourier space, see \S 
\ref{sect:torque_fourier}.


\section{Numerical results}
\label{sect:results}

 

\subsection{Numerical procedure.}  
\label{sect:numerics}

Having specified the gap density profile the solution for the 
azimuthal velocity perturbation $v$ is obtained by direct 
numerical integration of
the equations (\ref{eq:sigma}), (\ref{eq:tildeB}), 
(\ref{eq:pot})-(\ref{eq:Delta}); knowing
$v$ we find $u$ and $\delta\Sigma$ using equations 
(\ref{eq:u}), (\ref{eq:dS}). 

Numerical integration 
uses the same method as was employed in RP12 for 
uniform disks: we shoot numerical solutions from 
the origin and match them to the WKB outgoing 
waves far from the planet (see also \citet{KP93}). 
As shown in NGG, the exact homogeneous solution to
Equation (\ref{eq:v_eq}) for a homogenous disk, i.e. the 
parabolic cylinder functions, 
can be well approximated by the WKB outgoing waves
\ba
v(x\to\pm\infty)\sim\sqrt{\frac{2}{3k_{y}x}}e^{\pm i3k_{y}x^{2}/4h}
\label{eq:v_wkb}
\ea
when $x/h\gg (8/9)\left[1+(k_{y}h)^{2}\right]/(k_{y}h)^{2}$. 
Moreover, for this approximation to accurately represent an
inhomogeneous solution 
of Equation (\ref{eq:v_eq}) one requires $xk_y\gg1$, so 
that $\phi\ll1$ and $ \partial\phi/\partial x\ll1$.

To additionally account for the presence of a density gap in
the disk, we use the fact that for every profile
$\partial \Sigma_0/\partial x \to 0$ when $x$ is several 
times the characteristic width of the gap $\delta$ and, 
therefore, the asymptotic solution will still be
given by (\ref{eq:v_wkb}). All conditions together imply that
$x/h \gg \min\{1/(k_yh),1,\delta/h\}$, which is the criterion we 
use to match our numerical solutions to the outgoing waves. 

Our numerical calculations use a 4-th order Runge-Kutta
integrator with spatial resolution of $h/400$ and 
$x \in [-220h,220h]$ for 720 uniformly log-spaced values 
of $k_yh$ between 0.01 and 15, and potential softening 
length of $10^{-4}h$.

\begin{figure*}
\centering
\includegraphics[width=16cm,height=10cm]{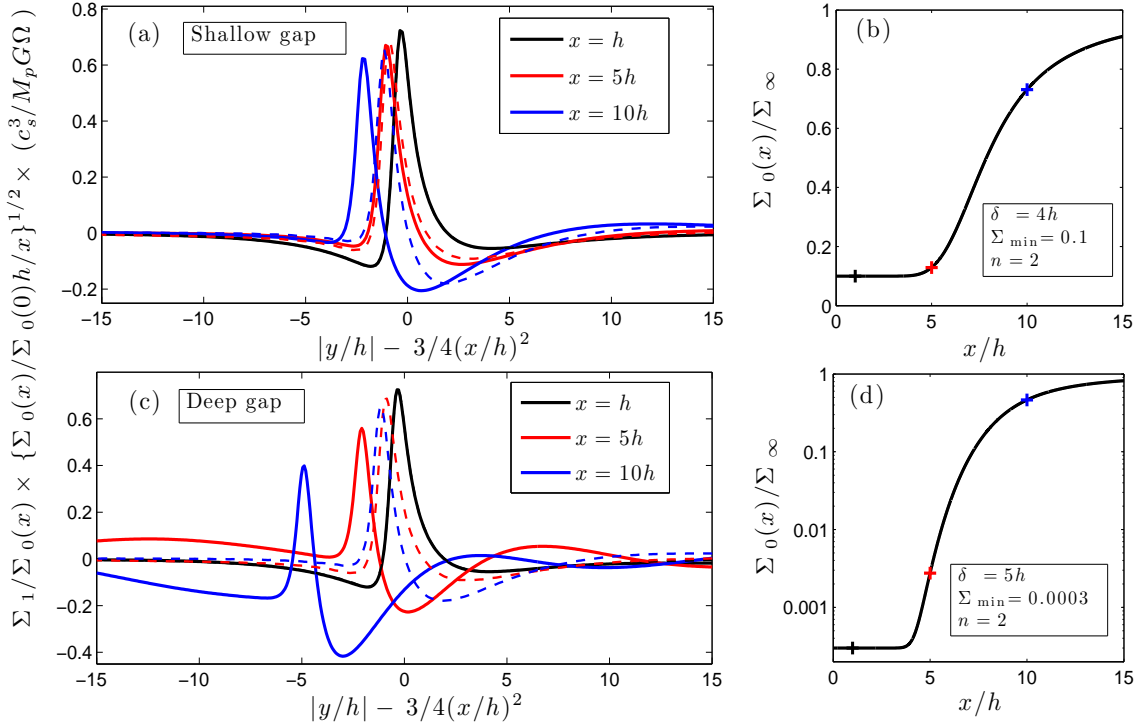}\\
\caption{Azimuthal cuts through the density wake in coordinate 
space showing the density perturbation $\Sigma_1(x,y)$
at different radial separations from the satellite 
($x=h,5h,10h$ as labeled), normalized by $[\Sigma_0(x)x]^{1/2}$,
and azimuthally shifted by $(3/4)x^2$ to facilitate comparison.
The solid lines in panels (a) and (c) show the results for the 
shallow gap profile (panel b; Eq. (\ref{eq:sigma_1}) with 
$\delta=4h$, $\Sigma_{\rm min}=0.1$, $n=2$) and 
deep profile (panel d; $\delta=5h$, 
$\Sigma_{\rm min}=3\times10^{-4}$, and $n=2$), 
respectively, while the dashed lines are results 
for a uniform density disk (normalized 
by  $[\Sigma_\infty x]^{1/2}$). 
Marks on the density profiles in right panels indicate
the radial position of the azimuthal cuts shown in left 
panels.
\\
\\
\label{fig:sigma_pert}
}
\end{figure*} 


\subsection{Density wake}
\label{sect:wake}

In Figure \ref{fig:sigma_pert} we show one of the outcomes 
of our calculations --- the behavior of the surface density 
perturbation $\Sigma_1(x,y)$ in physical space. We obtain it 
by first calculating $\delta\Sigma$ using Equation (\ref{eq:dS})
with the numerically determined $v$ and then performing the 
inverse Fourier transform. In Figure \ref{fig:sigma_pert}
azimuthal cuts of $\Sigma_1(x,y)$ at different values of $x$ 
are normalized by $(\Sigma_0(x)x)^{1/2}$ to ensure similar 
wake amplitude at different radial separations from the 
perturber; this scaling is expected\footnote{This expectation 
is only approximate since the angular momentum flux carried 
by the wake can vary substantially across the gap edge, see 
\S \ref{sect:T_F_H}. Nevertheless, the proposed scaling 
works quite well in practice, which is obvious from Figure 
\ref{fig:sigma_pert}.} from the conservation 
of the angular momentum flux carried by the density wave
\citep{GR01,rrr02a}.

In panel (a) we display our results for a shallow
gap with a minimum surface density at the 
satellite position of $0.1 \Sigma_\infty$ (corresponding 
density profile is shown in panel (b)). 
For comparison we also display 
the wake profiles for a uniform density disk.
Deep inside the gap, at $x=h$, where the $\Sigma_0$ 
is essentially constant in our model we find our
$\Sigma_1$ to exactly coincide with the uniform 
disk calculation, as expected (our normalization 
of $\Sigma_1$ takes care of the difference in amplitude).
However, further out from the planet the results start 
to differ. In particular, at $x=10h$ the azimuthal 
location of the wake in a non-uniform disk gets 
shifted by $\approx h$ relative to the uniform case,
but the differences in the shape of the density 
perturbation remain at the modest level.

Situation is a bit different for deep gaps, as 
illustrated in panel (c) where we plot azimuthal 
density cuts through the wake for a gap profile 
with $\Sigma_{\rm min}=0.0003 \Sigma_\infty$ (shown 
in panel (d)). Again, at $x=h$ density profiles 
coincide for both uniform and non-uniform cases.
However, at the gap edge (at $x=5h$) there are now 
substantial differences between the profiles not only in 
the wake position but also in shape. The differences 
become more pronounced as the wave propagates even further: 
at $x=10h$ the peak value of $\Sigma_1$ is shifted 
by $\sim 4h$ compared to the uniform case and the overall 
wake shape is considerably distorted. This is not 
surprising since the surface density eigenfunctions 
at the edge of the gap in a non-uniform disk are 
considerably distorted compared to their uniform disk 
analogues, see \S \ref{sect:torque_harmonics}.

Note that the amplitude of scaled $\Sigma_1$  
in Figure \ref{fig:sigma_pert} does not vary too much
with $x$.
Since the angular momentum flux conservation predicts
that $\Sigma_1\propto [\Sigma_0(x)x]^{1/2}$ it is clear
that the amplitude of $\Sigma_1$ {\it grows} as the wake
propagates out from the bottom of the gap. At the same time,
the relative density perturbation $\Sigma_1/\Sigma_0(x)$
scales as [$x/\Sigma_0(x)]^{1/2}$ and {\it decreases} as the
wake climbs up the density contrast at the gap edge. This
may have important implications for the wake damping, as 
discussed in \S \ref{sec:discussion}.


\section{Angular momentum transport: preliminaries}
\label{sect:ang_transport}


One of the key characteristics of the disk-satellite coupling
is the excitation torque density $dT/dx$ (or equivalently 
$dT/dr$) --- the amount of 
the angular momentum added by the satellite tide to the density 
wave per unit radial distance $x$. Following RP12 we express 
it via the imaginary part of the density perturbation 
$\Im(\delta\Sigma)$ as
\ba
\frac{dT}{dx}=\int\limits_{-\infty}^\infty
dy\delta\Sigma\frac{\partial\Phi}{\partial y}
&=&\int\limits_0^\infty \left(\frac{dT}{dx}\right)_{k_y} dk_y, \\
\left(\frac{dT}{dx}\right)_{k_y}&=&-4\pi k_y\phi\Im(\delta\Sigma),
\label{eq:dT_dx_k}
\ea
where $(dT/dx)_{k_y}$ is the torque density contribution due to mode
with wavenumber $k_y$. Integrated (one-sided) torque is then 
defined as $T(x)=\int_0^x 
\left(\partial T/\partial x' \right)dx'$.
Note that in the shearing sheet approximation there is no
net torque and $T(x)=-T(-x)$.
  
Integrated torque is closely related to the angular momentum flux 
(AMF) $F_H(x)$ --- the amount of angular momentum carried by the 
wake, which can be written as (RP12) 
\ba
F_H(x)&=&\int\limits_0^\infty F_{H,k_y}(x)dk_y,  \nonumber\\
F_{H,k_y}&=&4\pi\Sigma_0(x)\left[\Re(v)\Re(u)+\Im(v)\Im(u)\right],
\label{eq:F_H}
\ea
where $F_{H,k_y}$ is the angular momentum flux carried by a 
particular azimuthal mode of the wave. Angular momentum 
conservation demands the radial divergence of the AMF 
$dF_{H,k_y}/dx$ to be equal to $(dT/dx)_{k_y}$. RP12 have 
demonstrated this to be the case for a uniform disk and their
calculation can be trivially extended for an arbitrary non-uniform 
disk density profile. This is done (separately for each azimuthal 
mode) by writing $F_{H,k_y}$ in terms of $v$ using (\ref{eq:u}), 
differentiating with respect to $x$, and then rearranging 
the terms proportional to $v\partial v/ \partial x$ with the 
aid of equation (\ref{eq:v_eq}). As a result one indeed finds 
that $dF_{H,k_y}/dx=(dT/dx)_{k_y}$.


\subsection{Lindblad Resonance torque prescription}
\label{sect:GT80}

To derive the asymptotic torque density behavior 
(\ref{eq:dTdx_GT}) in physical space GT80 have explicitly 
assumed that the angular momentum associated with a 
particular low-order ($k_yh\lesssim 1$) harmonic of the 
perturbing potential gets added to the density wave 
exactly at the Lindblad resonance corresponding to that mode,
i.e $x_L(k_y)=2/(3k_y)$. 
\citet{dong11a} have extended this assumption to modes with
arbitrary values of $k_y$ to obtain a so-called 
\textit{Lindblad Resonance } (LR) prescription
for the angular momentum flux $F_H^{\rm LR}(x)$, which 
they used as a benchmark for comparing with their numerical 
results for uniform disks. This prescription is given by
the following formula \citep{dong11a}: 
\ba
F_H^{\rm LR}(x)&=&\frac{1}{h}\int\limits_{\mu_{min}(x)}^\infty d\mu 
\frac{F_H(\mu)}{F_H^{\rm WKB}(\mu)}F_H^{\rm WKB}(\mu), 
\label{eq:F_H_phys}\\
F_H^{\rm WKB}(\mu)&=&\frac{4}{3}\Sigma_\infty\mu^2
\left(\frac{G \mplanet}{c_s}\right)^2
\nonumber\\
&\times &\left[2K_0(2/3)+K_1(2/3)\right]^2,
\label{eq:flux_Fourier_theor}
\ea
where $F_H(\mu)/F_H^{\rm WKB}(\mu)$ is the ratio of the actual 
AMF contribution 
carried by a particular harmonic with $\mu\equiv k_yh$ far from the
perturber to the value of the same quantity obtained using WKB 
approximation. This ratio has been computed numerically by GT80
for arbitrary $\mu$; in the limit $\mu\to 0$ it tends to
unity. 

In equation (\ref{eq:F_H_phys}) $\mu_{min}(x)=(2/3)h/x$ is 
the minimum value of $k_yh$ for which the position of a 
corresponding Lindblad resonance (in a uniform disk) satisfies 
$x_L(k_y)<x$. Thus $F_H^{\rm LR}(x)$ represents the sum of the
AMFs (measured at $x\to \infty$) carried by all individual Fourier 
modes having their Lindblad resonances lying at $x_L(k_y)<x$.
Note that in computing $\mu_{min}(x)$ we do not use the
positions of Lindblad resonances accounting for disk 
nonuniformity (Equation \ref{eq:k_y}) --- we found this
prescription for  $x_L(k_y)$ to provide $F_H^{\rm LR}$
inconsistent with numerical results, see \S\ref{sect:dT_dx_full}.

The prescription (\ref{eq:F_H_phys}) is formulated for a 
disk with constant density $\Sigma_\infty$. We extend 
it to disks with gaps by first defining a {\it Lindblad 
resonance torque density} in a uniform disk as 
$dT^{\rm LR}(x)/dx|_u\equiv dF_H^{\rm LR}(x)/dx$,
and then using this torque density in equation 
(\ref{eq:dTdx_nu}): 
\ba
\frac{dT^{\rm LR}(x)}{dx}=\frac{\Sigma_0(x)}{\Sigma_\infty}\cdot
\frac{dF_H^{\rm LR}(x)}{dx}.
\label{eq:dT_dx_LR}
\ea
Note that for $|x|\gtrsim h$ one has $\mu_{min}\lesssim 1$ and 
$dT^{\rm LR}(x)/dx|_u$ reduces to the standard GT80 
expression (\ref{eq:dTdx_GT}). Most studies of tidal coupling 
in non-uniform disks have used this asymptotic form of the 
prescription (\ref{eq:dT_dx_LR}), i.e. equation (\ref{eq:dTdx_nu}) 
with $dT/dr|_u$ given by equation (\ref{eq:dTdx_GT}), to 
characterize the torque density; use of equation (\ref{eq:F_H_phys}) 
simply allows us to extend this prescription to arbitrary 
values of $x$.

The prescription (\ref{eq:F_H_phys}) is equivalent to assuming 
that each potential harmonic exerts torque only in the immediate 
vicinity of its Lindblad resonance, a notion that has been 
shown by RP12 to be incorrect in the shearing sheet 
approximation even for $\mu\lesssim 1$. In reality Lindblad 
resonances have finite width, which leads to the nontrivial 
interference between them, resulting in the negative torque 
density phenomenon at $|x|\gtrsim 3.2 h$ 
(Dong et al. 2011a; RP12)
not captured by the GT80 analysis. Despite this failure of 
the assumed discreteness of the Lindblad resonance, we still 
use $F_H^{\rm LR}(x)$ in this work to compare with our new results  
as it provides an interesting reference point.


\section{Torque behavior in Fourier space}
\label{sect:torque_fourier}


We start our investigation of the angular momentum 
exchange between the disk and the perturber by exploring the 
torque behavior for individual Fourier harmonics. We explore 
both the spatial structure of the torque density due
to individual modes (\S \ref{sect:torque_harmonics}), 
as well as their contribution to the full torque far from the 
perturber (\S \ref{sec:fourier}).


\subsection{Spatial structure of $\left(dT/dx\right)_{k_y}$}
\label{sect:torque_harmonics}

\begin{figure*}
\centering
\includegraphics[width=18.3cm,height=12cm]{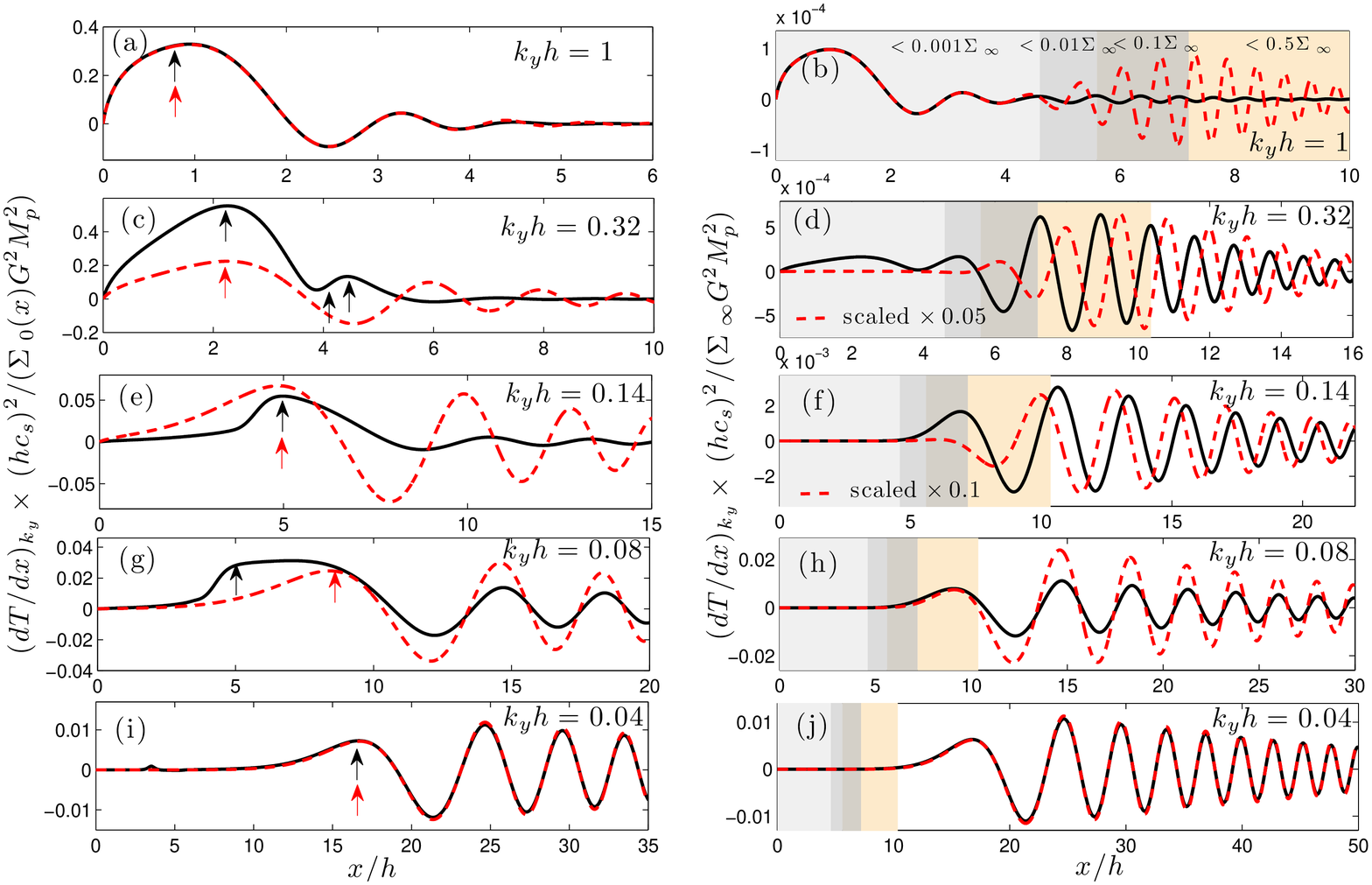}\\
\caption{ \textit{Left panels}: torque density per unit  
surface density $\Sigma_0(x)$ for the non-uniform disk 
(black solid lines) and per $\Sigma_\infty$ for the uniform 
density case (red dashed lines) produced by modes of a given $k_y$. 
\textit{Right panels}: same curves but multiplied by the 
non-uniform density profile $\Sigma(x)/\Sigma_\infty$ (i.e. black solid
lines now show pure non-uniform disk torque density $(dT/dx)_{k_y}$).
Gap model given by Eq. (\ref{eq:sigma_1}) with parameters 
$\delta=5h$, $\Sigma_{\rm min}=3\times10^{-4}$, and $n=2$.
 Black and red arrows in the left panel indicate
the position of the Lindblad resonances in non-uniform and uniform 
disks, respectively. 
In panels (d) and (f) we scale the results for the uniform case by 0.05
and 0.1, respectively, to facilitate comparison. Shaded regions
in right panels indicate (from left to right) the regions where
$\Sigma_0(x)$ is less than $0.001\times\Sigma_{\infty}$,
$0.01\times\Sigma_{\infty}$, $0.1\times\Sigma_{\infty}$, 
and $0.5\times\Sigma_{\infty}$.
\\
\\
\label{fig:dT_dx_k}
}
\end{figure*}

In Figure \ref{fig:dT_dx_k} we show the torque density 
for individual Fourier harmonics $(dT/dx)_{k_y}(x)$ 
computed using equation (\ref{eq:dT_dx_k}), at different 
radial separations $x$. These calculations assume our 
fiducial density profile (\ref{eq:sigma_1}) with parameters 
$\delta=5h$, $\Sigma_{\rm min}=3\times10^{-4}$, and $n=2$. 
We show both the torque density per unit local surface density  
(i.e. $(dT/dx)_{k_y}(x)$ normalized by $\Sigma_0(x)$; left 
panels) as well as the pure $(dT/dx)_{k_y}(x)$ (divided by
constant $\Sigma_\infty$, right panels). 
We also display $(dT/dx)_{k_y}|_u$ computed for a uniform disk
(see Figure 3 of RP12), normalizing it by $\Sigma_\infty$ 
in the left panels (which yields the torque density per unit
local surface density in a homogeneous disk) and also 
multiplying by  $\Sigma_0(x)/\Sigma_\infty^2$ (to incorporate 
the local density correction) in the right panels. In all 
cases torque density is properly normalized to make it
dimensionless. In the left panels of this Figure we indicate 
the position of the Lindblad resonances with arrows, both 
for an inhomogeneous (see equation [\ref{eq:k_y}]) 
and a uniform disks.

In panels (a) and (b), we plot $(dT/dx)_{k_y}(x)$  for $k_y h=1$. 
The corresponding Lindblad resonance lies at $x/h=2/3$, in the flat 
bottom of the gap and is thus the same for both uniform and 
fully non-uniform disks. Moreover, the waveform inside the gap 
perfectly matches that of the uniform disk since the profile 
has constant density there. Note that the torque density is 
not sharply localized at the Lindblad resonance position 
$x_L(k_y)$ but is extended over a reasonably broad radial
range around $x_L$, with quite pronounced (even though damping) 
oscillations of $(dT/dx)_{k_y}(x)$ clearly visible even for
$x\gg x_L$. This additionally confirms the finite and 
non-negligible width of resonances previously pointed 
out by \citet{arti93} and RP12.

At $x>4h$, where $\Sigma_0$ starts to increase, we observe that 
the torque density 
decreases its amplitude quite abruptly compared to the uniform 
disk calculation, see panel (b). By looking at the shaded 
regions one can notice that the two calculations start 
differing at $\Sigma_0\sim 10^{-3}\Sigma_\infty$ where
the density gradients become significant. This damping of the 
amplitude compared to the homogeneous disk case is the 
consequence of incorporating the disk non-uniformity 
in the fluid equations, also clearly visible for other values of 
$k_yh$ at the gap edge (see below).

In panels (c) and (d) we show the torque density for $k_y h=0.32$ mode, 
for which there are multiple Lindblad resonances at $x/h=2.1$, 
3.8, and 4.5, see Figure \ref{fig:lindblad}  
(the uniform disk has, of course, just a single one at 
$x/h=2/(3 k_y h)\approx 2.1$).
These are indicated with arrows in panel (c)
and it can be seen from both panels that the extra 
resonances located at $x\sim 4h$ give rise to positive
torque density contribution and overall significant phase shift 
compared to the uniform disk case. Also, the amplitude of the 
torque density oscillations far from the resonances is greatly 
reduced in the self-consistent calculation of $(dT/dx)_{k_y}(x)$ 
so that in panel (d) we even multiply the uniform disk curve by 
$0.05$ for it to have an amplitude comparable to
that of the fully non-uniform disk calculation.

The mode with $k_yh\approx 0.14$ shown in Figure \ref{fig:dT_dx_k}e,f
has a single Lindblad resonance at $x_L\approx 5h$, which coincides
with the resonance position in a uniform disk (see Figure 
\ref{fig:lindblad}). Despite this coincidence the waveforms 
for the uniform and non-uniform disk calculations are very 
different from each other both in terms of shape and the 
overall amplitude (note that the uniform disk curve in panel 
(f) has been multiplied by $0.1$). 

The mode with $k_yh=0.08$ shown in Figure \ref{fig:dT_dx_k}g,h
has its (single) Lindblad resonance at $x\approx 5.1h$, 
considerably shifted inward from the corresponding uniform 
disk resonance position at $x=8.3h$. The torque density normalized 
by $\Sigma_0(x)$ (see panel (g)) has a rather broad spatial 
distribution extending from $x\sim 5h$ so $x\sim 8h$ before starting 
to oscillate at an amplitude reduced roughly by a factor of $2$
compared to the uniform disk calculation. 

Finally, in panels (i) and (j) we show the mode $k_yh=0.04$ 
for which $x_L\approx 16h$ lies where the density gradients are 
negligible. As a result, calculations in a uniform and nonuniform
disk cases  perfectly coincide, demonstrating that far from gap 
torque excitation is well described by the uniform disk theory 
of RP12.

To summarize, the spatial distribution of the torque density 
for individual Fourier harmonics can be substantially modified 
in non-uniform disks in regions where the density gradients 
are large, compared to the case of a constant density disk. 
The differences arise both because of the displacement of
the Lindblad resonances and because of the different mathematical 
structure of the fluid equations in the non-uniform disk case.


\subsection{Angular momentum flux in Fourier space}
\label{sec:fourier}

\begin{figure}
\centering
\includegraphics[width=8.8cm,height=12cm]{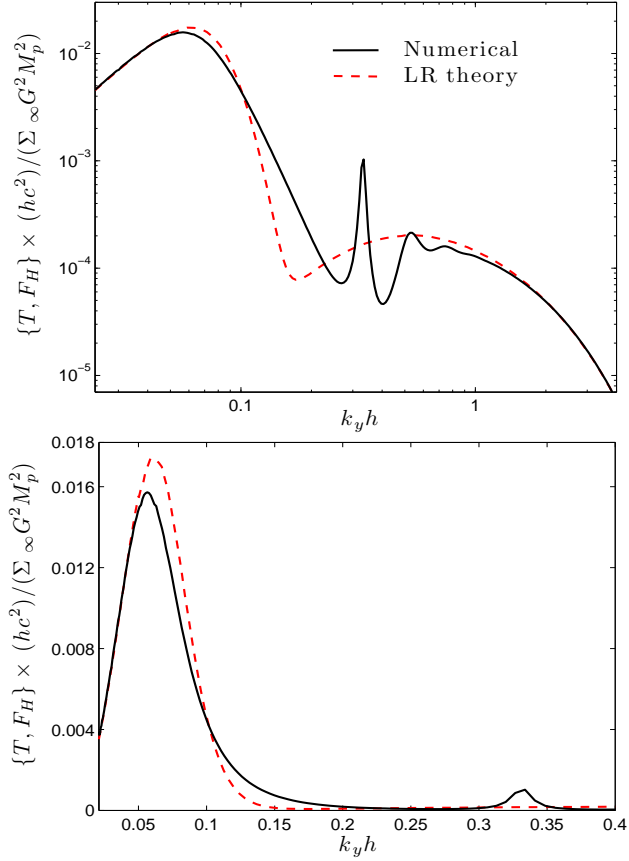}
\caption{Angular momentum flux (or integrated torque) far from 
the planet ($x\to\infty$) in Fourier space (logarithmic 
coordinates at the top, linear at the bottom) for a gap model 
given by Eq. (\ref{eq:sigma_1}) with  
$\delta=5h$, $\Sigma_{\rm min}=3\times10^{-4}$, and $n=2$.
The black solid line indicates the results from equation (\ref{eq:T_k})
and the red dashed line results from the LR theory
described in \S \ref{sec:fourier}.
}
\label{fig:T_k}
\end{figure} 

We now study the behavior of the AMF as a function of $k_y$
for $x\to \infty$. Since, as mentioned in \S 
\ref{sect:ang_transport}, $F_{H,k_y}(x)$ is identical (up 
to a constant offset) to the integrated torque 
we look at the behavior of 
\ba
T(k_y)=\int\limits_0^\infty \left(\frac{dT}{dx}\right)_{k_y} dx
\label{eq:T_k}
\ea 
in Fourier space, where $(dT/dx)_{k_y}$ is given by our 
numerical calculations.

As a point of comparison we also use the variation of 
$F_H(k_y)\equiv F_{H,k_y}(x\to\infty)$ with $k_y$ in a 
homogeneous disk (previously computed by GT80) which we
additionally multiply by $\Sigma_0(x_L(k_y))$ to account 
for the disk non-uniformity (where we take 
$x_L(k_y)=2/(3k_y)$). This re-scaling implies that each 
mode carries the AMF from a uniform disk corrected
by the local density at the position where the excitation 
of this mode occurs in a uniform disk. This procedure is 
a version of the LR theory described in \S \ref{sect:GT80} 
in Fourier space, as it incorporates the two major 
ingredients from that prescription: the localized mode 
excitations at a Lindblad resonance and the AMF computed 
for a uniform density disk.

In Figure \ref{fig:T_k}, we plot the torque from our calculations 
and the AMF from the LR theory for our fiducial model with
$\delta=5h$, $\Sigma_{\rm min}=3\times10^{-4}$, and $n=2$.
The first thing we notice is that both expressions coincide in the 
limits of long wavelengths $k_yh<0.05$ and short wavelengths
$k_yh>2$. This is expected to happen since in both limits the 
satellite excites waves in the flat parts of the disk --- far 
outside the gap for $k_yh<0.05$ and deep inside the gap, in 
the flat part of the density profile for $k_yh>2$. For that 
reason the positions of the resonances and the waveforms 
coincide in our non-uniform calculation and in a uniform disk.
But in the intermediate range of $k_y$ there are appreciable 
differences between the LR theory and our calculations, mainly 
due to effect of the shifted Lindblad resonances as we discuss 
next.

First, for $k_yh\sim 0.05-0.1$ Figure \ref{fig:lindblad} 
demonstrates that Lindblad resonances are shifted towards 
the perturber and the excitation occurs in {\it lower} density 
region of the disk compared to the case when resonances lie 
at $x=2/(3k_y)$ (as appropriate for homogeneous disk). Because
of this reduction of the background surface density the torque 
produced by modes in this range of $k_y$ ends up being lower 
than the AMF computed according to the LR theory (which assumes
$x_L=2/(3k_y)$), even though the coupling to the perturbation 
potential is stronger at smaller $|x|$. Conversely, modes with 
$k_yh\sim 0.1-0.2$ have their resonances shifted {\it outwards}, 
into the higher density part of the gap profile, which explains 
their higher values of $T(k_y)$ compared to the LR theory AMF.

At higher values of $k_y$ there is a point where resonances split 
and take place at multiple locations, as discussed
in \S \ref{sect:lindblad}. According to Figure \ref{fig:lindblad} 
this is the case for $k_yh\gtrsim 0.2$, which explains the intricate 
shape of $T(k_y)$ for these modes --- a peak at $k_yh=0.2-0.3$
where the Lindblad resonances are split in three (see Figure 
\ref{fig:dT_dx_k}c,d for illustration of the torque density 
behavior for these modes). The innermost resonance almost 
coincides with that in a uniform disk, but the other two are 
located further out, in a higher density region of the gap, 
explaining the spike of $T(k_y)$ for these values of 
$k_y$. 

Complicated structure of the $T(k_y)$ curve around the peak at 
$k_yh\sim 0.32$ does not have a straightforward  explanation and 
is probably related to the intricate behavior of the torque 
density when the excitation is taking place at multiple locations 
at the gap edge.


\section{Torque behavior in physical space}
\label{sect:torque_physical}


We now explore the behavior of the torque integrated over all
Fourier modes in physical space. We look at the spatial 
distribution of the torque density in \S \ref{sect:dT_dx_full} 
and of the integrated torque $T(x)$ in \S \ref{sect:T_F_H}.


\subsection{Torque density}
\label{sect:dT_dx_full}

\begin{figure}
\centering
\includegraphics[width=9cm,height=13cm]{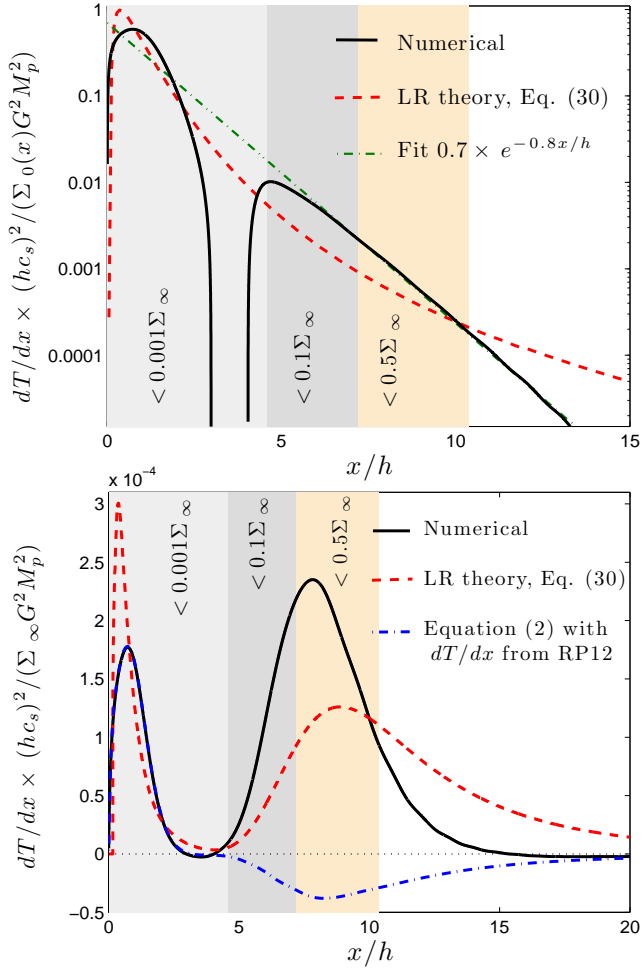}
\caption{Torque density $dT/dx$ obtained self-consistently for
a non-uniform disk using our numerical
results (solid black lines) and that from the LR
prescription (\S \ref{sect:GT80}; dashed red lines) for
 a gap model given by Eq. (\ref{eq:sigma_1}) with parameters 
$\delta=5h$, $\Sigma_{\rm min}=3\times10^{-4}$, and $n=2$.
In the \textit{upper panel} we normalize by the surface density 
$\Sigma_0(x)$ and add a numerical fit to show the
\textit{exponential} fall-off of $dT/dx$ outside the gap.
In the \textit{lower panel} we normalize by the density at
infinity $\Sigma_\infty$ and also display $dT/dx$ computed 
using the prescription in Eq. (\ref{eq:dTdx_nu}) with
the uniform disk torque density properly accounting for the 
negative excitation torque density phenomenon (RP12). The latter
curve demonstrates the conceptual failure of a simple 
prescription (\ref{eq:dTdx_nu}).
\label{fig:dT_per_sigma}
}
\end{figure}

In the top panel of Figure \ref{fig:dT_per_sigma}, 
we plot the torque density $dT/dx$ (integrated over all 
$k_y$, see equation [\ref{eq:dT_dx_k}]) per unit of 
the local surface density $\Sigma_0(x)$ for our fiducial model
and make a comparison with the LR theory behavior 
given by equation (\ref{eq:dT_dx_LR}). In the lower panel
of Figure \ref{fig:dT_per_sigma} we directly compare 
$dT/dx$ and $dT^{\rm LR}/dx$. One can clearly 
see a significant discrepancy between the two.

The disagreement at $|x|\lesssim 5h$, where the surface 
density is close to constant (the bottom part of the gap) 
is expected based on the RP12 results --- one can see
that our $dT/dx$ is negative for $x\sim (3-4)h$ which is 
just the manifestation of the negative torque density 
phenomenon \citep{dong11a} not captured by the $dT^{\rm LR}/dx$.
Also, the overall shape of $dT/dx$ and $dT^{\rm LR}/dx$
is different for $|x|\lesssim 3h$ --- again well known 
from the uniform disk studies of \citet{dong11a} and RP12.

The discrepancies at $|x|\gtrsim 5h$ where the density 
gradients are important can only be understood by fully 
accounting for the disk non-uniformity.
According to GT80 for $|x|\gtrsim h$ the LR theory 
predicts $\Sigma_0^{-1}dT/dx\approx 2.5  (h/x)^4$ 
(in this section we refer to the torque density in units of 
$\Sigma_\infty G^2M_p^2/(hc_s)^2$), see equation (\ref{eq:dTdx_GT})
and \S\ref{sect:GT80}. However, our self-consistent nonuniform 
disk calculations suggest that the torque density decays 
\textit{exponentially} $\propto\exp{(-0.8x/h)}$ for 
$|x|\gtrsim 5h$, as shown in the upper panel of Figure 
\ref{fig:dT_per_sigma}. This result casts serious doubt on all 
torque prescriptions used in the literature that adopt the GT80 
functional form to account for the tidal torque. The exponential 
fall-off is clearly related to the presence of strong density 
gradients at the gap edge but we could not find a simple 
qualitative explanation for this particular form of the 
torque decay. We just note here that this behavior of $dT/dx$
has nothing to do with the presence of the exponential term
in equation (\ref{eq:sigma_1}), which is introduced only to
guarantee the flat density profile in the bottom part of the 
gap. We provide more details on the exponential decay of 
$dT/dx$ in Appendix \ref{sec:fit}.

Apart from the discrepancy in the functional form, the 
self-consistent specific torque density is also more 
concentrated towards the perturber at the gap edge, and 
has a larger amplitude there. This effect can be understood 
from our previous analysis of the shifted Lindblad resonances 
that tend to accumulate at the gap edge (see Figures 
\ref{fig:lindblad} and \ref{fig:dT_dx_k}), increasing excitation 
torque density in this region and reducing it outside of it, 
at $|x|\gtrsim 10h$. This effect has not been taken into 
account in previous studies but it does drastically change 
the overall shape of the torque density. 

We tried to improve the performance of the LR theory 
(\S \ref{sect:GT80}) by assigning torque contributions
of individual modes not to $x_L=2/(3k_y)$ (which is appropriate 
only for a uniform disk), but to $x_L(k_y)$ given by the 
self-consistent calculation in a non-uniform disk, see 
equation  (\ref{eq:k_y}). Unfortunately, the agreement with 
our fully self-consistent calculation has gotten even worse
(for this reason we did not pursue this idea further),
which suggests that properly accounting for the resonance 
overlap is also very important for getting the correct 
spatial distribution of the torque.

Far enough from the gap edge, where $\Sigma_0(x)$ flattens out, 
the excitation torque density becomes negative, which is a direct
manifestation of the negative torque density phenomenon 
\citep{dong11a}. For the adopted gap profile this happens at 
$x\approx 15h$, and $dT/dx$
stays negative out to infinity and agrees with
the asymptotic behavior $-0.63(h/x)^4$ derived in RP12. 

Additionally, in the bottom panel of Figure 
\ref{fig:dT_per_sigma} (dot-dashed curve) 
we plot $dT/dx$ calculated according 
to the simplified prescription (\ref{eq:dTdx_nu}) but using the  
correct excitation torque density for the uniform disk 
$dT/dx|_u$ calculated in RP12 instead of the GT80 prescription 
(\ref{eq:dTdx_GT}). One can see that this calculation fails
miserably compared to the correct $dT/dx$: it predicts negative 
excitation torque density for $x>3.2h$, which is the direct 
consequence of the negative torque density phenomenon 
in the uniform disks. Because of this, as we will later see in Figure 
\ref{fig:dT_dx_AMF}, the use of this prescription results in
the {\it negative} integrated torque deposited in the density 
wave by the perturber, which cannot be true in a real disk.

The point of this last exercise was to illustrate the failure of the
simple prescription (\ref{eq:dTdx_nu}), which yields unphysical 
results even when it uses physically motivated ingredients --- the
spatial behavior of $dT/dx|_u$ derived in RP12 as opposed to
the asymptotic formula (\ref{eq:dTdx_GT}).


\subsection{Integrated torque and AMF}
\label{sect:T_F_H}

\begin{figure}
\centering
\includegraphics[width=8.8cm,height=8cm]{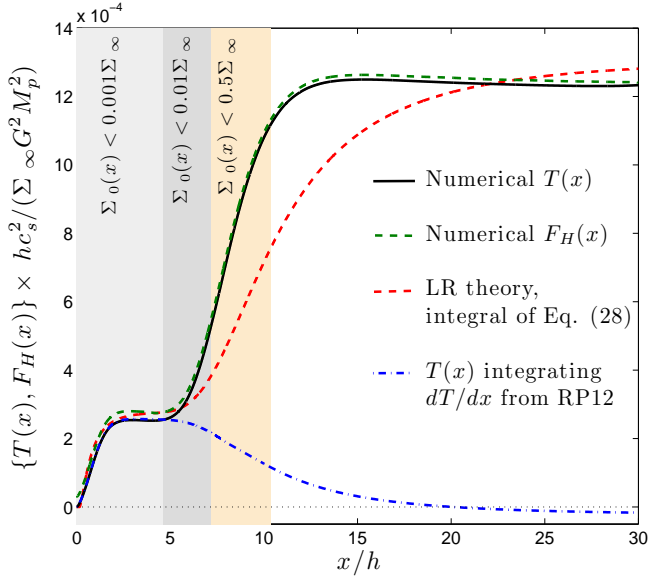}
\caption{Integrated torque $T(x)$ from our numerical 
results (solid black line), the LR theory (dashed red line),
and the integration of the torque density from uniform density disk
(RP12) using the prescription (\ref{eq:dTdx_nu})
(dot-dashed blue line). Dashed green line displays the AMF 
$F_H(x)$ obtained from Equation (\ref{eq:F_H}) and, as expected, 
agrees with $T(x)$. Note that the RP12 prescription 
for $dT/dx|_u$ used in combination with equation (\ref{eq:dTdx_nu})
results in {\it negative} integrated torque as $x\to\infty$, which
is unphysical (see \S \ref{sect:dT_dx_full} for details).
Results are for a gap model (\ref{eq:sigma_1}) with  
$\delta=5h$, $\Sigma_{\rm min}=3\times10^{-4}$, and $n=2$.}
\label{fig:dT_dx_AMF}
\end{figure}

In Figure \ref{fig:dT_dx_AMF}, we plot the 
integrated torque $T(x)$ for our fiducial model as well as 
the AMF $F_H(x)$ calculated using equation (\ref{eq:F_H}). 
The two have small relative vertical offset but are otherwise 
identical, as expected from the discussion in \S 
\ref{sect:ang_transport}. We also display $T^{\rm LR}(x)$ obtained
by integrating $dT^{\rm LR}/dx$ given by equation (\ref{eq:dT_dx_LR})
from zero to $x$. 

All these expressions almost coincide within the gap. This 
happens because in the flat-bottomed gap different methods
of computing integrated torque should yield the same result: 
a total torque (or AMF) of $\approx 0.93$ (GT80) in units 
of the plot times $\Sigma_{\rm min}=3\times 10^{-4}$. This
equivalence of the full torque calculations in the Fourier 
space ($T^{\rm LR}(x)$) and physical space ($T(x)$) has been
previously mentioned in RP12.

At the gap edge, for $|x|\gtrsim 5h$, one finds that $T(x)$
increases faster than $T^{\rm LR}(x)$, which is a direct 
consequence of the torque density $dT/dx$ in a self-consistent 
non-uniform disk calculation being rather concentrated towards 
the perturber, see \S \ref{sect:dT_dx_full}. Quite remarkably, 
despite this difference at the gap edge the accumulated torque 
at infinity nearly coincides with that given by the simple theoretical 
prescription $T^{\rm LR}(x)$: our self-consistent calculation 
yields full torque $\approx 1.2\times 10^{-3}$, while the 
the LR theory predicts $\approx 1.3\times 10^{-3}$ (see Figure 
\ref{fig:dT_dx_AMF} at $x=30h$ for reference). This means that 
the {\it total} torque exerted on the disk (but not the spatial 
distribution of the torque density!)
is well described by the simple theoretical prescription
(essentially $T^{\rm LR}(x)$ based on asymptotic scaling 
[\ref{eq:dTdx_GT}]) that has been broadly used in the 
literature. 

We do not have a fully satisfactory explanation for this 
coincidence except for the following observation. In the 
lower panel of Figure \ref{fig:T_k} we display the torque 
in Fourier space (on linear scale) for $k_y h<0.4$. 
There it can be seen that our non-uniform disk calculation 
generate almost the same amount of torque as the theoretical
prescription (integrals under the curves are roughly the same),
meaning that the torque deficit with respect to the LR theory 
visible for $k_yh\lesssim 0.1$ is almost exactly compensated 
by the torque excess for $k_yh\gtrsim 0.1$. This means that 
the effect of resonances, which are being shifted to lower 
density regions resulting in lower torque contribution (smaller
$k_yh$) is closely counterbalanced by the torque enhancement
due to the resonances displaced to higher density region 
(larger $k_yh$). This is an interesting result, which only 
weakly depends on the exact gap profile as we empirically 
show next.


\section{Effect of varying gap shape parameters}
\label{sect:params}


So far, we showed the results obtained for a specific
gap density profile given by equation (\ref{eq:sigma_1}) 
with a certain set of parameters, namely $\delta=5h$, 
$\Sigma_{\rm min}=3\times10^{-4}$, and $n=2$. It would 
of course be much better to discuss torque structure 
not for an arbitrary model of the gap profile 
but for the one which is realized in nature. However, 
the calculation of such profile must self-consistently couple the 
calculation of the torque excitation (such as the one
done here) with the description of the wave damping, and
this is beyond the scope of this work.
However, we can still explore the general {\it trends}
in the torque behavior as the various properties of the
gap (its depth, width, etc.) are varied. This is what 
we do now.

\begin{figure*}
\centering
\includegraphics[width=18.5cm,height=11.5cm]{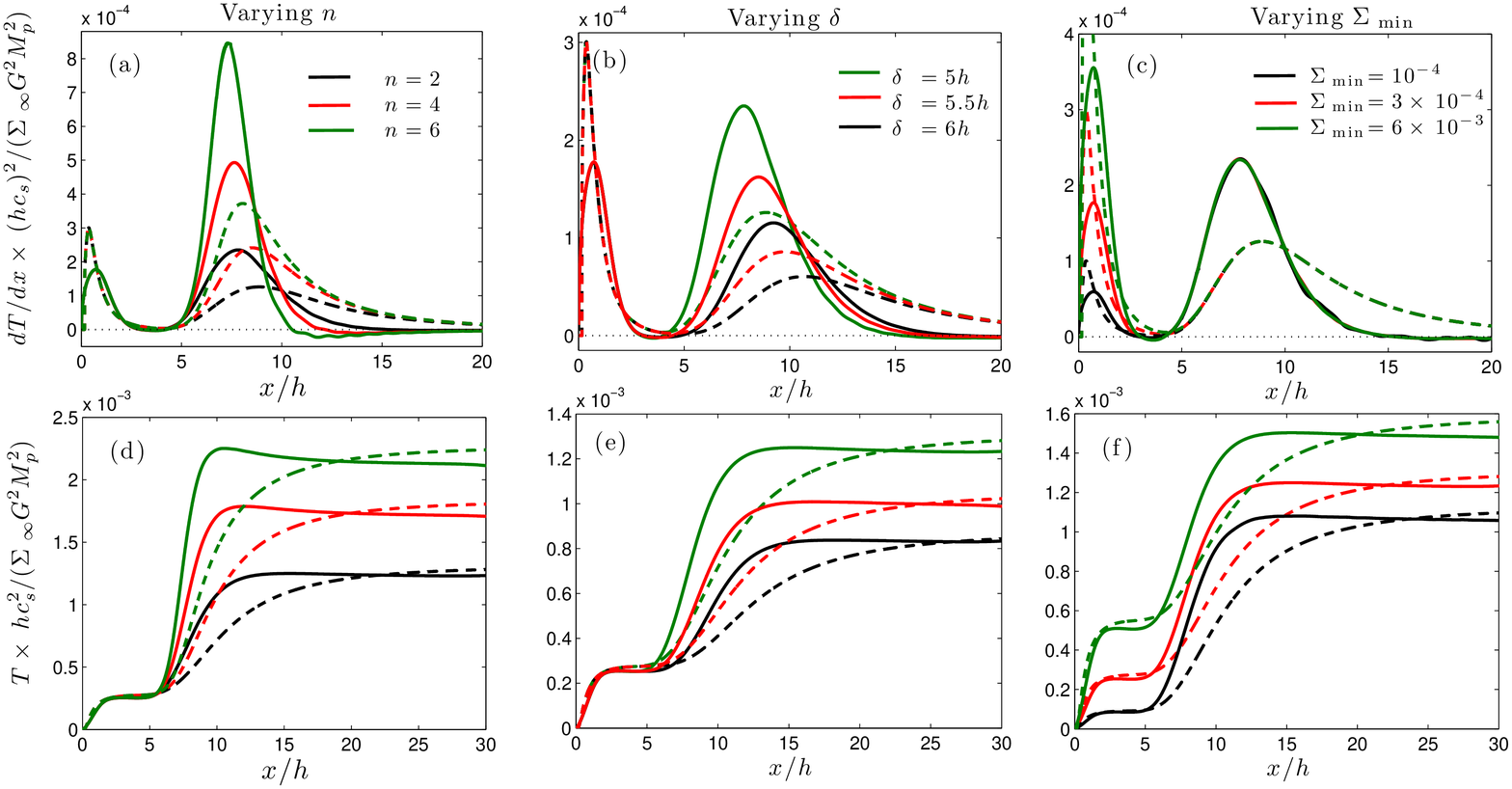}\\
\caption{Torque density (upper panels a, c, and e) and 
integrated torque (lower panels b, d, and f) for
 a gap model given by Eq. (\ref{eq:sigma_1}) varying the 
three parameters  $n$, $\delta$, and $\Sigma_{\rm min}$ 
one at a time with respect to our fiducial choice 
 $\delta=5h$, $\Sigma_{\rm min}=3\times10^{-4}$, and $n=2$.
In panel (a) and (d) we vary $n$, fixing $\delta$, $\Sigma_{\rm min}$,
in panel (b) and (e) we vary  $\delta$, and 
in panel (c) and (f) we vary  $\Sigma_{\rm min}$.
Dashed line indicates the results from the LR 
theory (\S \ref{sect:GT80}).
\\
\\
}
\label{fig:vary_all}
\end{figure*}


\subsection{Steepness of the gap edge (variation of $n$)}

As discussed in \S \ref{sect:models} and shown in 
Figure \ref{fig:sigma_profiles}, the parameter $n$ controls 
the steepness of the density gradients at the gap edge. In panels 
(a) and (d) of Figure \ref{fig:vary_all} we plot 
the excitation torque density and the integrated torque, respectively, 
for $n=2, 4, 6$, while fixing  $\delta=5h$ and 
$\Sigma_{\rm min}=3\times 10^{-4}$. 

As $n$ is increased the gradients become larger and, 
therefore, the amplitude of the variation of $\kappa^2$ around
the gap edge becomes larger. This leads to higher 
resonance accumulation 
closer to the perturber, and $\Sigma_0$ gets higher there as well.
As a result, the torque density gets more concentrated towards 
the gap center with increasing $n$, see Figure \ref{fig:vary_all}a:  
for $n=6$ almost all the torque is excited within the range 
$5h\lesssim x\lesssim 10h$, while for $n=2$ this region extends 
out to $x\sim 15h$.

The integrated torque also increases with $n$ since 
closer to the gap center excitation by the perturber is more
effective, depositing more angular momentum into the density 
waves. We see from Figure \ref{fig:vary_all}d that the density 
profile with $n=2$ results in a total torque of 
$\sim 1.2\times10^{-3}$ in units of 
$(GM_p)^2\Sigma_\infty/hc_s^2$, while using $n=6$ we obtain 
$\sim 2.1\times10^{-3}$.

This Figure also demonstrates that for all values of $n$
the LR theory predicts more extended torque density distribution
than we find in self-consistent calculations, see \S 
\ref{sect:dT_dx_full}. Nevertheless, the integrated torque 
at $|x|\to\infty$ is still well predicted by the LR theory, 
the point we previously made in \S \ref{sect:T_F_H}. The 
largest relative difference of $\sim 13 \%$ is found 
for $n=6$, for which we obtain the integrated torque of 
$\sim 2.1\times10^{-3}$, while the LR prescription gives 
$\sim 2.4\times10^{-3}$.


\subsection{Width of the gap (variation of $\delta$)}
\label{sect:gapwidth}

In Figure \ref{fig:vary_all}b,e we vary the width of the gap 
by slightly changing $\delta$, but fixing 
$\Sigma_{\rm min}=3\times10^{-4}$ and $n=2$.
As expected, for narrower gaps the torque density peaks 
closer to the planet and the integrated torque reaches 
higher values, simply because the strength of the tidal 
coupling is higher closer to the perturber. 

Figure \ref{fig:vary_all}e shows that the integrated torque 
predicted by the LR theory tends to better match the 
self-consistent calculation as the gap width is increased. 
Also, the dependence of the $T(x\to\infty)$ on $\delta$ is 
not linear: decreasing $\delta$ from $6h$ to $5.5h$ results 
in $\sim 14 \%$ increase of the integrated torque, while 
going from $5.5h$ to $5h$ results in the increase of 
$\sim 25 \%$. Using the fact that the LR theory gives a 
reasonably good approximation to the integrated torque 
behavior, and adopting the prescription (\ref{eq:dTdx_nu}) 
with $dT/dr|_u$ given by equation (\ref{eq:dTdx_GT}) one 
can estimate that the total torque excited outside the 
density gap scales as $T(x>\delta)\propto(\delta/h)^3$.
We verified this simple scaling by calculating 
integrated torque for density profiles with $\delta=
(4-8)h$. This behavior is in agreement with \citet{LP84} 
who present the same scaling for the one-sided torque.


\subsection{Gap depth (variation of $\Sigma_{\rm min}$)}
\label{sec:vary_sigma}

In Figure \ref{fig:vary_all}c,f we vary the dimensionless 
gap depth $\Sigma_{\rm min}$ from $10^{-4}$ to $10^{-3}$, 
while fixing $n=2$ and $\delta=5h$. One can see that changing 
$\Sigma_{\rm min}$ only affects the torque density and the full 
torque accumulated inside the gap (in a linear fashion); 
outside the gap (for $x>4h$) torque structure remains 
unchanged. 

As $\Sigma_{\rm min}$ is increased there is a point at which 
the integrated torque starts being dominated by the excitation
inside the gap, in the immediate vicinity of the perturber. 
One can easily predict when this happens. For a flat-bottomed 
gap profile like the one we consider in this work the torque
accumulated inside the gap is $T(x<\delta)\sim 0.9 
\Sigma_{\rm min} \Sigma_\infty G^2M_p^2/hc_s^2$ (RP12 and GT80)
if the gap width $\delta$ is larger than about $2h$ --- the 
extent or the region where most of the torque excitation occurs in 
a uniform disk. This exceeds the torque $T(x>\delta)$ excited
outside the gap for
\ba
\Sigma_{\rm min}\gtrsim \frac{T(x>\delta)hc_s^2}
{0.9\Sigma_\infty G^2M_p^2}\approx 2.8h^3
\int\limits_\delta^\infty\frac{\Sigma_0(x)}{x^4}dx,
\label{eq:sigma_min_min}
\ea
since the value of $T(x>\delta)$ is well approximated by the 
LR theory using the prescription (\ref{eq:dTdx_nu}) 
with $dT/dr|_u$ given by equation (\ref{eq:dTdx_GT}).
In the case shown in Figure \ref{fig:vary_all}c,f the critical 
value of $\Sigma_{\rm min}$ is $\approx 10^{-2}$, i.e. the 
integrated torque in this case is dominated by the excitation 
at the edge of the disk only when the density inside
the gap is depleted below $1\%$ of 
$\Sigma_\infty$.

Finally, after trying different combination of parameters
for the density profile, we observe that the torque density
always decays {\it exponentially} outside the gap in regions
where the density gradients are important, as discussed
in \S \ref{sect:dT_dx_full}. In Appendix B we construct an 
empirical fit for the torque density in this region 
represented by equation (\ref{eq:fit}).


\section{Discussion}
\label{sec:discussion}


The main result of this work is a new, self-consistent 
calculation of the torque excitation by a massive perturber 
in a non-uniform disk. We clearly show that the existing 
torque density prescriptions 
\citep{LP86,trilling98,armi-nata02,armitage02}, 
majority of which are based on the GT80 asymptotic 
result (\ref{eq:dTdx_GT}), provide inadequate 
description of the spatial distribution of $dT/dr$ in 
disks with gaps. 

In particular, we find that for deep gaps the torque 
excitation is more concentrated towards the perturber
than the standard prescription (\ref{eq:dTdx_nu})
with $dT/dr|_u$ from equation (\ref{eq:dTdx_GT}) would 
predict. Also, in regions where density 
gradients are non-zero we find the fall-off of the torque 
density to be \textit{exponential} as opposed to the $x^{-4}$ 
dependence that is assumed in all previous works
(see \cite{GT80,LP79,LP84,LP86} and references
therein). Moreover, far from the gap edge, where the 
density flattens out, we recover the negative torque 
density phenomenon previously found by \citet{dong11a} and 
RP12 in a uniform disk. 

There are three main reasons for these differences:
\begin{enumerate}

\item Lindblad resonances in non-uniform disks are
shifted with respect to their constant density disk analogues, 
and often occur at multiple locations (\S \ref{sect:lindblad}).

\item Spatial behavior of the eigenmodes of the perturbed
fluid variables is different in a non-uniform disk 
(\S \ref{sect:torque_harmonics}).

\item Interference of different Lindblad resonances important
for both uniform and non-uniform disks is usually not accounted
for by the LR-type prescriptions (\S \ref{sect:GT80}).

\end{enumerate}

One can try to improve the existing semi-analytical torque
prescriptions. Previously there have been some attempts to 
incorporate the (weak) disk non-uniformity in the LR 
prescription for $dT/dr$ by accounting for only the 
resonance shift \citep{menou04,ward97,matsumura07}. 
We checked the performance of this procedure in \S 
\ref{sect:dT_dx_full} but found that it does not improve 
(at least for strong density variations) the agreement 
with the self-consistent calculation. And the other two 
issues listed above are clearly very difficult to account 
for in any simple way. 

Unfortunately, an accurate calculation of $dT/dr$ for
an arbitrary $\Sigma_0(r)$ profile, which is necessary for
studies of the coupled evolution of the perturber and the 
surrounding disk \citep{LP86,trilling98,ivanov99,armitage02,
chang10}, is rather computationally-intensive as 
outlined in \S \ref{sect:numerics}. In this regard it would 
be nice to have a simple analytical fit to the actual 
excitation torque density behavior and in Appendix \ref{sec:fit}
we provide a simple one-parameter formula (\ref{eq:fit}) 
for $dT/dr$ for exactly this purpose. Even though there is some 
ambiguity in choosing the proper fit (it depends on a single 
parameter $\alpha$, which is somewhat different for different 
gap profiles) this approach provides a basis for finding 
better semi-analytical representations of the $dT/dr$
behavior useful for long-term numerical calculations of
the disk-satellite interaction.

Despite the different predictions for the torque density 
$dT/dr$ we found that the 
LR theory outlined in \S \ref{sect:lindblad} does produce a 
reasonable description of the {\it total integrated} torque, 
within $\sim 20\%$ of that found 
in a self-consistent calculation for different values of
the gap density profile, see \S \ref{sect:params}. 
Previously \citet{LP84} carried out a numerical torque 
calculation for a polytropic disk truncated by a massive 
perturber located at the distance $\delta$ from the disk 
edge, and found that the integrated torque scales as 
$\sim \delta^{-3}$. This is in agreement with our results 
described in \S \ref{sect:gapwidth} and supports the 
observation that the LR theory provides reasonably accurate
description of the full one-sided torque exerted on the disk.

Based on this remarkable result one may decide 
that the orbital evolution of the perturber should be 
insensitive to the discrepancy in the torque density 
calculation described above, simply 
because the variation of the perturber's angular momentum 
depends on the full integrated torque it exerts on the 
disk. This, however, is not true since the 
integrated torque does sensitively depend on the gap density 
profile (see \S \ref{sect:params}), and the latter is a 
function of the torque density distribution in the disk. 
Thus, $dT/dr$ is in fact involved (even though indirectly) 
in determining the rate of migration of the perturber.
This makes any statements regarding the Type II migration
speed in protoplanetary disks and the rate of SMBH inspiral 
due to interaction with the circumbinary disk 
dependent on our understanding of the torque density in 
non-uniform disks, which our work aims to provide.


\subsection{Nonlinearity of the density wave}
\label{sect:nonlinearity}

We noticed in \S \ref{sect:wake} that the conservation of 
the angular momentum forces the relative amplitude of the 
density wake $\Sigma_1/\Sigma_0$ to {\it decrease} as the 
wave climbs up the density gradient at the gap edge. 
Interestingly, even though the normalized $\Sigma_1/\Sigma_0$ 
does not vary much as the wave travels through the gap edge, 
the angular momentum density varies a lot; this is because the 
latter is sensitive to the overall wake {\it shape} and not just its 
amplitude. 

The variation of the relative density perturbation 
has interesting implications for the strength of 
the density waves outside the gap. In particular,
from the fact that the normalized wake amplitude in
Figure \ref{fig:sigma_pert} does not vary much with $x$ 
it follows that the wave nonlinearity measured by the ratio 
$\Sigma_1/\Sigma_0\propto\left(x/\Sigma_0(x)\right)^{1/2}$ 
should decrease by a factor of 
$\sim \left[(\delta/h)/\Sigma_{\rm min}\right]^{1/2}$ between
the bottom of the gap and the part of the disk just outside
of the gap, where $\Sigma_0\sim \Sigma_\infty$. For clean 
gaps with small $\Sigma_{\rm min}$ this factor can be 
quite significant, $\approx 0.2$ for a gap with the width of 
$\delta=5h$ and density contrast at the bottom 
$\Sigma_{\rm min}=10^{-2}$ (this factor is $\sim 0.05$ for the 
situation shown in Figure \ref{fig:sigma_pert}c). 
Thus, even if the density wave starts out highly 
nonlinear ($\Sigma_1/\Sigma_0\gtrsim 1$) 
inside the gap near the planet, it may become linear upon
propagating out of the gap. The initial wave nonlinearity
in the flat part of the density profile near the planet is
$\Sigma_1/\Sigma_0\sim M_p/M_{\rm th}$, where 
$M_{\rm th}\equiv c_s^3/(\Omega G)$ is the characteristic 
planetary mass at which the Hill radius becomes comparable
to the disk scale height. This mass is about 10M$_\oplus$
at 1 AU and about 60M$_\oplus$ at 10 AU. Thus, a Jupiter mass
planet at 10 AU would generate a highly nonlinear wave with
$\Sigma_1/\Sigma_0\sim 5$ at the bottom of the gap but the 
nonlinearity should drop to $\Sigma_1/\Sigma_0\sim 1$
outside the gap for $\delta=5h$ and 
$\Sigma_{\rm min}=10^{-2}$, even if we neglect the wave 
damping (not captured by our linear theory). 
This demonstrates that even rather massive 
planets capable of opening gaps may still have their density 
waves in the linear regime outside the gap, depending mainly 
on the density contrast $\Sigma_{\rm min}$.

This point is rather important for the possibility of the 
nonlinear wave damping. \citet{GR01} have studied the
nonlinear density wave evolution in a uniform disk 
and demonstrated that shock formation resulting 
from nonlinear effects can lead to effective damping of the 
wave. This study has been extended by \citet{rrr02a} to
include the possibility of radial variation of $\Sigma_0$.
Our linear results suggest, in agreement with this latter work, 
that the nonlinear wake distortion (which ultimately results 
in shock formation) should slow down as the wave gets out
of the gap, shock formation gets postponed and the nonlinear
dissipation becomes less effective at transferring wave angular
momentum to the disk material.

For massive planets the wave might shock close to the planet,
while it is still propagating through the roughly uniform 
bottom part of the gap profile: \citet{GR01} estimate the 
shocking length of just $2h$ for a $2M_\oplus$ planet at 1 
AU, and it is even smaller for more massive planets. Then
the efficient density wave dissipation will start inside 
the gap but then will appreciably slow down as the wave 
climbs out of the gap (even though the shock will still 
persist). This complication should certainly affect the
picture of the nonlinear wave damping and may produce 
important feedback for the self-consistent calculations of 
the gap density profile.


\subsection{Astrophysical implications}
\label{sect:implications}

Theory of disk-satellite interaction has been extensively
used for understanding the interaction of protoplanetary 
disks with embedded planets and the evolution of SMBH binaries
surrounded by circumbinary accretion disks, among other 
astrophysical settings. Our extensions of this theory have 
direct impact on our understanding of these systems.

In particular, our finding of the torque density being
quite concentrated near the perturber suggests that 
regardless of the details of the wave damping, the angular 
momentum carried by the density waves should be deposited 
in the disk {\it closer to the perturber} than the 
conventional theory would suggest. This certainly facilitates
the formation of density gaps or cavities since for the same 
mass of the perturber a narrower annulus of the disk needs
to be cleaned by the tide. This may make Type II migration 
possible for lower mass planets than was previously thought.

To fully understand the impact of our results on the Type II 
migration and SMBH inspiral one needs to know a {\it 
self-consistent gap shape}, which can be computed only when 
the wave damping mechanism is specified. In this regard we note 
that the majority of studies of the gap or cavity opening 
simply ignore the issue of wave damping process and take
$dT/dr|_d = dT/dr$ \citep{LP86,chang10}, i.e. assume immediate 
deposition of the excitation torque into the disk.
However, the validity of this approximation has never been 
demonstrated, and calculations that do properly take 
wave damping into account \citep{ward97,rrr02} result in a rather 
different picture of the gap opening. Our results on the 
wave amplitude evolution and its implications for the 
nonlinear wave damping in \S \ref{sect:wake}, 
\ref{sect:nonlinearity} should be useful for understanding 
the spatial distribution of $dT/dr|_d$ and calculating 
the gap shape (and orbital evolution of the perturber) 
fully self-consistently.

Our present work is cast in terms of the shearing sheet 
approximation, which limits the direct application of its 
results to systems in which gaps are narrow compared
to the semi-major axis of the perturber. This is the case 
for e.g. planets in protoplanetary disks 
\citep{LP86,takeuchi96} and SMBH binaries with extreme mass 
ratio surrounded by an accretion disk (so-called extreme 
mass-ratio inspirals (EMRIs)) 
\citep{armi-nata02,chang10,kocsis11}.

SMBHs binaries of similar mass do not fall in this category 
since they are known to clear out a large and clean 
cavity in circumbinary gaseous disk, with inner radius 
comparable to the binary separation
(e.g.  \citet{mac08}, \citet{cuadra09}, \citet{shi11}, 
\citet{farris11}). However, even though the shearing sheet 
approximation breaks down in these systems, our results 
can still be used to gain qualitative insight into the 
dynamics of cavity clearing. In particular, concentration
of the torque density towards the perturber means that cavity 
should be opened by lower mass ratio SMBH than was thought 
before.

Also, our results can help understand the excitation 
torque density patterns derived from simulations, most of which 
\citep{mac08,cuadra09,shi11,farris11,roedig12} agree that outside the cavity
$dT/dr$ exhibits complicated oscillatory pattern (some of these
studies include additional physics such as MHD or general 
relativity). This behavior was not recognized before and 
some earlier studies (e.g. \citet{armi-nata02}) tried modeling
numerically derived $dT/dr$ in the form similar to the GT80's 
result (\ref{eq:dTdx_GT}) but with the amplitude reduced by 
a factor of $\sim 10^{-2}$. Such low amplitude most likely 
results from averaging actual spatially rapidly oscillating 
pattern of $dT/dr$ and then applying the $\propto |r-r_p|^{-4}$
scaling to reproduce the total integrated torque in the 
simulations. Clearly, usage of this prescription is likely to 
lead to misleading results \citep{liu2010,chang10} in 
modeling gap shape.

\citet{mac08} suggested that the oscillatory behavior of the 
torque distribution in their simulations is consistent with 
forcing at the 2:3 ($m=2$) outer Lindblad resonance and tried 
fitting linear waveforms from \citet{meyer87} to spatial variation 
of the perturbed fluid variables. Even though they were able 
to reproduce qualitatively the oscillatory behavior of the 
numerical $dT/dr$ the amplitude and phase of the torque 
density were substantially different 
between the numerical calculation and the analytical waverform,
qualitatively consistent with the Lindblad resonance shift and 
reduction of perturbation amplitude in our calculations, see
Figure 4f. 
This is certainly not surprising since not only were the \citet{meyer87}
waveforms derived for a uniform disk, they are also valid
only in the close vicinity of the Lindblad resonance. The latter
limitation was removed in RP12 who came up with a fully global 
analytical approximation for the waveforms, but again only for 
a uniform disk case. 
Our present work is free from these
constraints since it provides a way of
computing $(dT/dr)_{k_y}$ for arbitrary disk surface density 
profile at any distance from the resonance, see 
\S \ref{sect:torque_harmonics}. 
Future extensions of our 
approach to full cylindrical geometry will be directly 
applicable to the astrophysical systems such as SMBH 
binaries in which broad gaps or cavities are typical.


\section{Summary}
\label{sec:conclusions}


In this work we have carried out a linear study of the 
tidal disk-satellite interaction in the shearing sheet 
approximation assuming the disk to have a non-uniform density
structure to model the effect of density gaps around the 
perturber. The disk non-uniformity is self-consistently
incorporated in the fluid equations, including the modification
of the disk rotation curve due to pressure gradients, which
are especially prominent at the gap edges. This allows
accurate calculation of the primary fluid variables
such as the density perturbation and torque distribution 
in both physical and Fourier space.

Because of the rotation curve modification the Lindblad 
resonances get shifted towards the gap edge, often occurring
at two or three locations on one side of the perturber. 
Because of this and the terms in fluid equations related to 
density gradients the waveforms of the modes excitated close 
to the gap edge are considerably modified compared to those 
obtained in a uniform density disk. We find the torque 
density in physical space to be more concentrated towards
the perturber and different from the uniform disk theory
predictions typically by a factor of several in this region. 
At least for the particular gap profiles considered in 
this work we find that the torque density at the gap edge 
drops exponentially with the distance from the perturber 
(as opposed to the power law scaling expected from uniform 
disk theory). In parts of the disk where the density is 
roughly uniform we observe the negative torque density 
phenomenon in agreement with the results of \citet{dong11a}
and RP12. Despite these differences in the torque density 
distribution the total angular momentum flux driven by the 
perturber through the disk seems to agree with the existing 
prescriptions at the level of $\sim 10\%$. The relative 
perturbation of the surface density is shown to go down 
in amplitude as the density wave propagates out of 
the gap, which has important implications for the 
nonlinear wave evolution and damping.

These results suggest that the process of gap opening 
in protoplanetary disks and gas clearing around SMBH 
binaries would be more efficient than the current 
theory predicts.
Our revision of the torque density prescription should have 
important effect on the self-consistent calculation of the gap 
density profile and, consequently, on the orbital evolution 
of the perturber (Type II migration speed for planets and
SMBH inspiral rate). Future extension of our results to the 
full cylindrical geometry will allow direct applications 
of the non-uniform disk theory to systems with wide gaps 
or cavities.

\acknowledgements{The financial support for this work is provided by 
the Sloan Foundation, NASA grant NNX08AH87G, NSF grant 
AST-0908269, and CONICYT Biccentenial 
Becas-Chile fellowship awarded to CP.\\\\}


\appendix
\section{Shifted Lindlblad resonances in the shearing-sheet approximation}
\label{res_shift}

The angular frequency of the non-uniform isothermal disk in full 
cylindrical geometry is given by 
\begin{eqnarray}
\Omega^{2}(r)=\Omega^{2}_{K}(r)+\frac{c_s^{2}}{r}
\frac{\partial \ln \Sigma_{0}}{\partial r},
\end{eqnarray}
where $\Omega_{K}$ in the regular Keplerian frequency.
The modified epicyclic frequency is then given by 
\begin{eqnarray}
\kappa^{2} = 4\Omega^{2} + 2r\Omega \frac{\partial \Omega}{\partial r}=  
\Omega_{K}^{2}(r) +c_s^{2} \frac{\partial^{2} \ln \Sigma_{0}}{\partial r^{2}} +
\frac{3c_s^{2}}{r} \frac{\partial \ln \Sigma_{0}}{\partial r} .
\end{eqnarray}
The locations $r_L$ at which the Lindblad resonance condition 
$m^2\left[\Omega(r_L)-\Omega_p\right]^2=\kappa^2(r_L)$ (here 
$\Omega_p\equiv\Omega_K(r_p)$ is the pattern speed of the 
gravitational perturbation and $m=k_yr_p$ is the wavenumber) are 
given by the following relation:
\begin{eqnarray}
m^2 \left[\sqrt{\Omega_{K}^2(r_L)+\frac{c_s^{2}}{r}
\frac{\partial \ln \Sigma_{0}}{\partial r}\Big|_{r_L}} - 
\Omega_{p}\right]^2 = \Omega_{K}^{2}(r_L) + c_s^{2} 
\left(\frac{\partial^{2}\ln\Sigma_{0}}{\partial r^{2}} +
\frac{3}{r} \frac{\partial \ln \Sigma_{0}}{\partial r}\right) \Bigg|_{r_L}.
\label{eq:m_interm}
\end{eqnarray}
Expressing $m$ via $k_y$ we find the following relation between 
$k_y$ and corresponding $r_L$, which does not make any approximations 
and is valid in full cylindrical geometry:
\ba
k_y(r_L)=\pm\frac{\kappa(r_L)}{\Omega_p r_p}
\left[\sqrt{\left(\frac{r_p}{r_L}\right)^3+\frac{h^{2}}{r}
\frac{\partial \ln \Sigma_{0}}{\partial r}\Big|_{r_L}}-1
\right]^{-1},~
\kappa^2(r_L)=\Omega_p^2\left[\left(\frac{r_p}{r_L}\right)^3+
h^{2}\left(\frac{\partial^{2}\ln\Sigma_{0}}{\partial r^{2}} +
\frac{3}{r} \frac{\partial \ln \Sigma_{0}}{\partial r}\right)\Big|_{r_L}
\right].
\ea

In the shearing sheet approximation one takes $r_L=r_p+x_L$, 
expands $r_p/r_L$ to linear order in $x_L/r_p$ and then takes 
the limit $r_p\to\infty$. As a result one recovers the resonance 
condition in the form (\ref{eq:k_y}) with $\kappa$
given by equation (\ref{eq:kappa}).


\section{A one-parameter torque density fit}
\label{sec:fit}

\begin{figure}
\centering
\includegraphics[width=16cm,height=7.5cm]{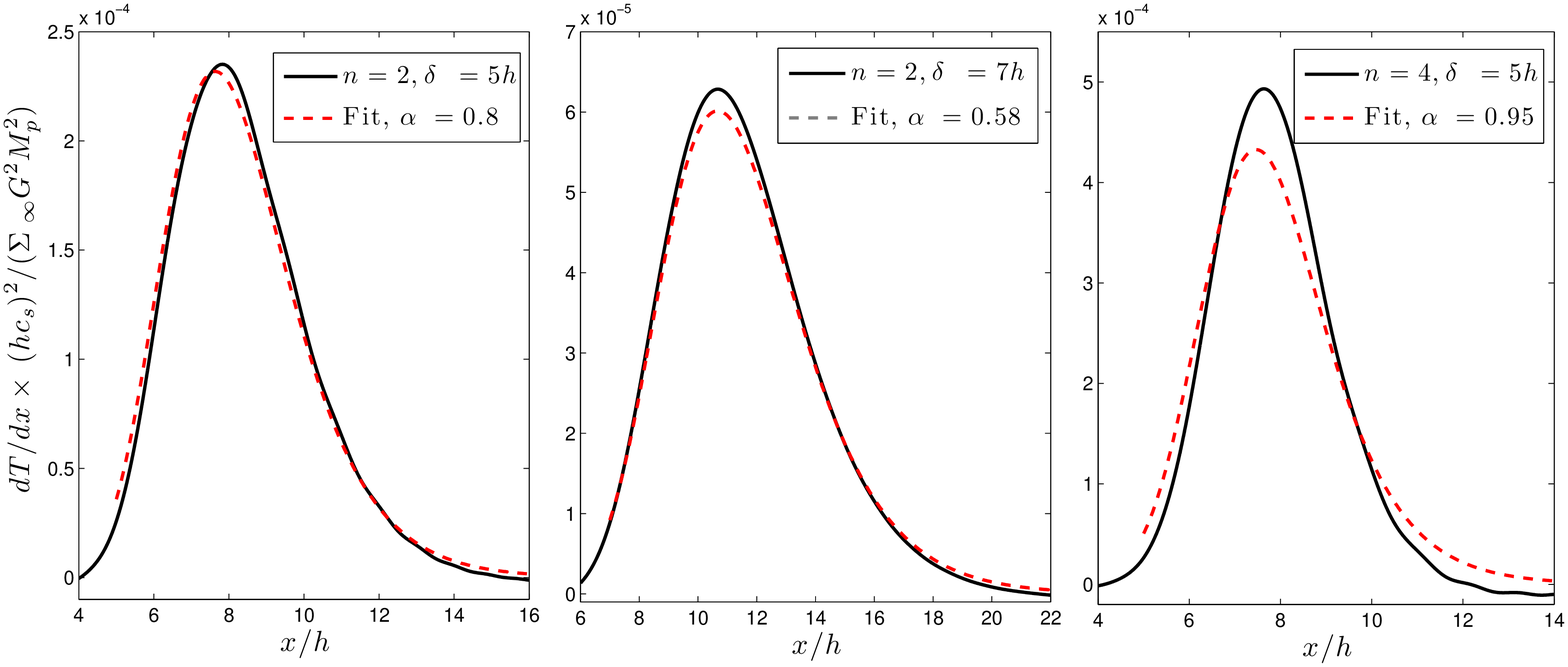}
\caption{Torque density outside the gap
for three different models (\ref{eq:sigma_1}) with 
$\Sigma_{\rm min}=3\times 10^{-4}$, and different $n$ 
and $\delta$ as specified in each panel.
Black curves show our numerical results while the red 
dashed curves show the fit (\ref{eq:fit}) with the 
input parameter $\alpha$ indicated in each panel.
}
\label{fig:fit}
\end{figure} 

Given the numerical challenges involved in the calculation
of the torque density in a general non-uniform disk we 
try to provide an empirical fit for $dT/dx$. Our procedure
is based on the following logic.

For every gap model considered in this work we find that the torque
density per unit surface density $\Sigma_0$ decays exponentially 
outside the gap until it becomes negative because of the
negative torque density phenomenon (RP12).
In particular, for our fiducial model the fall-off is 
described reasonably well by $\exp(-0.8x/h)$  for $x/h>5$ 
as shown in Figure \ref{fig:dT_per_sigma}.
Also, as discussed in \S \ref{sect:T_F_H}, \ref{sect:params} 
the full integrated torque $T(\infty)$ coincides
within $\sim 10\%$ with the prediction based on simple 
GT80 prescription (\ref{eq:dTdx_GT}) and (\ref{eq:dT_dx_LR}). 
We use this property to get a normalization to the 
exponential fall-off outside the gap. Putting these 
two ideas together we come up with the following 
fit for the torque density outside the gap ($x>\delta$):
\ba
\frac{dT}{dx}=C(\alpha) {\rm sign}(x) 
\frac{(GM_p)^2\Sigma_0(x)}{\Omega^2h^4}
e^{-\alpha x/h}, ~~~
C(\alpha)=2.5\frac{\int_{\delta}^{\infty}\Sigma_0(x)/(x/h)^4dx}
{\int_{\delta}^{\infty}\Sigma_0(x)e^{-\alpha x/h}dx},
\label{eq:fit}
\ea
where $\alpha$ is a dimensionless free parameter
that controls the exponential fall-off and takes value 
in the range $\sim 0.5-1$ for all our numerical calculations 
as we show next.

In Figure (\ref{fig:fit}), we show how this fit performs for 
different gap profiles as we vary the width
or steepness of the density profiles (the torque density
outside the gap does not depend on its density contrast 
$\Sigma_{\rm min}$, see \S \ref{sec:vary_sigma}). 
In the left panel, we consider our fiducial model and 
the fitting formula using $\alpha=0.8$. 
Our prescription fits the numerical result pretty well 
from $5h\lesssim x\lesssim 15h$, while beyond $15h$ 
the torque density becomes negative and the correct 
asymptotic behavior is given by the analytical expression 
$-0.63\Sigma_0(x)/x^4$, see RP12 (in proper units). 

In the middle panel, we increase the width of the gap
by taking $\delta=7h$ and show a fit with $\alpha=0.58$,
which works well for $7h\lesssim x\lesssim 22h$, before 
torque density changes sign.

In the right panel, we increase the steepness of the gap profile
by using $n=4$ and show the best fit (\ref{eq:fit}),
which now requires $\alpha=0.95$. In this case, the fit is worse 
than in the previous examples with relative errors up to 
$\lesssim 20\%$ at the peak of the torque density, but it 
still does a better job than the asymptotic expression from
GT80.

Unfortunately, at the moment we cannot predict a 
value of the parameter $\alpha$ featured in equation 
(\ref{eq:fit}) for a particular gap profile. But the results 
shown in Figure \ref{fig:fit} suggest that the torque density
decay is faster and the fits demands higher values of $\alpha$  
for boxier density profiles (i.e. for higher $n$) and for
narrower gaps (i.e. for lower $\delta$).

\end{document}